\documentclass[12pt,oneside]{amsbook}
\usepackage{epsfig}
\numberwithin{equation}{chapter}
\numberwithin{figure}{chapter}

\theoremstyle{plain} 

\usepackage{mathrsfs}
\usepackage[left=3cm,right=3cm,top=5cm,bottom=2cm]{geometry}
\theoremstyle{definition}

\theoremstyle{remark}

\begin{document}

\frontmatter 
\title[]{$\mu$-$\tau$ Symmetry and Leptogenesis in the Minimal Seesaw Model}
\author{Daniel Wegman Ostrosky}
\address{Department of Physics\\CINVESTAV-IPN\\Mexico City\\Mexico}
\date{November 16, 2007}
\maketitle

\newpage

\vspace*{4cm} 

\bigskip\bigskip\bigskip\bigskip\bigskip\bigskip\bigskip\bigskip\bigskip\bigskip
\textit{I dedicate this thesis to my parents; to my mother for introducing me to the world of science, and for inherit me the love for it. And to my father, who thought me to follow my dreams and to work hard to archive them.}

\newpage

\begin{center}
\textbf{Resumen}
\end{center}
{\small Los valores medidos para los \'angulos de mezcla en oscilaciones de neutrinos sugieren la existencia de una simetr\'ia de intercambio de sabor entre neutrinos $\mu$  y  $\tau$.  Usando esta simetr\'ia se analiza el modelo seesaw m\'inimo para masas de neutrinos, en donde se diagonaliz\'o la masa de Majorana, y se demuestra que el modelo soporta a lo m\'as 3 fases de violaci\'on CP y 5 masas reales a altas energ\'ias. Sin embargo a bajas energ\'ias solo se manifiestan 4 par\'ametros de masa y una fase CP de Majorana. Por lo tanto, usando los valores medidos experimentalmente de las diferencias cuadradas de masas, los \'angulos de mezcla y la jerarqu\'ia, se pueden determinar algunos par\'ametros del modelo pero no todos. Se propone el uso del par\'ametro de asimetr\'ia bari\'onica del universo debido a leptog\'enesis para determinar una fase m\'as del modelo.
Para finalizar se uso una jerarqu\'ia normal de las masas de los neutrinos de mano derecha para hacer  una aproximaci\'on, que permite reconstruir completamentela matriz de masas de neutrinos de mano izquierda. En especial se determina el valor de $m_{ee}$ el cual se podr\'a comparar con los resultados del doble decaimiento beta sin neutrinos.}

\vspace{.4cm}
\begin{center}
\textbf{Abstract}
\end{center}
{\small The measured values for the mixture angles in neutrino oscillations suggest the existence of a symmetry of interchange of flavor between neutrinos $ \mu$ and $ \tau$. Using this symmetry we analyzed the minimal seesaw model for neutrino masses, where the  Majorana mass was diagonalized, and it is  demonstrated that the model supports at most 3 CP violation phases and 5 real masses at high energies. Nevertheless, at low energies, only 4 parameters of mass and one relative Majorana CP phase remain. Therefore using the experimental values of the masses square differences, the mixture angles and the hierarchy,  we can determine some parameters of the model but not all. Also we propose the use of the parameter of baryonic asymmetry of the universe due to leptogenesis to determine one more phase of the model. Finally we used a normal hierarchy for the masses of the right handed neutrinos to make an approximation, that allowed us to completely reconstruct the mass matrix for left handed neutrinos. In special the value of $m_ {ee}$ is determined which can be compared with the results of the neutrinoless double decay beta.}

\newpage       			
\chapter*{Acknowledgments}

First of all I would like to thank my thesis advisor, Dr. Abdel P\'erez Lorenzana, his help has been invaluable. 

Secondly, to my family, for always being there. We've had some rough times, but they have always encourage me to go forward.

To the friends I made over the master and to the ones from before, their help, support, and especially friendship, had help me the get where I am now.

Finally, I also would like to thank CONACYT for the financial support given to me during the masters degree and to CINVESTAV and all my teachers there, for being a key ingredient in my education. 

\tableofcontents



\mainmatter 

\newcounter{pnumb}                     	
\setcounter{pnumb}{\value{page}}       	
\mainmatter                            	
\setcounter{page}{\value{pnumb}}       	

    			
\chapter*{Introduction}

The universe is an open question, still, all the time we are learning more about it. 100 years ago we didn't even knew how matter is composed, scientists believed that  the atoms were indivisible and though the fundamental components of all that exist. Now we know much more, atoms are composed of a nucleus and electrons surrounding it, the nucleus is a combination of protons and neutrons, and they are made of quarks. So, at the moment we believe that quarks and electrons are fundamental particles. But these are not the only particles we have discovered; a complete particle zoo is now part of our theories that try to explain our existence.  

\vspace{.3cm}

The Standard Model has been a victory of modern physics. It describes the strong, weak, and electromagnetic interactions, and their synthesis down to millifermis. Most of its predictions have been found by almost half a century of experiments and all the particles that it predicts have been seen experimentally; including leptons, neutrinos and bosons, with the only exception of the Higgs boson (which is probably to be found in following experiments like LHC).

\vspace{.3cm}

Still it is believed to be incomplete. First it doesn't include the forth fundamental interaction, the gravity. Second it has too many parameters that have to be measured experimentally, and with the technology we have, we still can't measure some of them. Third, it doesn't provide any understanding for the dark matter and energy in the universe. And finally (and the motivation for this thesis), whereas the Standard Model predicts that neutrinos don't have mass; today we know this isn't true, making it a handicap model, with a need for improvement.

\vspace{.3cm}

There had been many theories and models that try to explain the neutrino mass and what it implies about physics beyond the Standard Model. In order for neutrino to have mass one possibility is to introduce a right handed neutrino in the model (since they don't exist in the Standard Model), in this thesis we have chosen a minimal scenario were the right handed neutrino is a singlet (instead of, for example a part of a doublet joint to a right handed charged lepton). In second place we have used the seesaw model to explain why right handed neutrinos have not been detected directly, this have been a successful model that predicts a light left handed neutral lepton (the standard neutrino), and a heavy right handed neutral lepton (which is referred in many places as the right handed neutrino), both of this are Majorana particles, being their own antiparticles.

\vspace{.3cm}

The problem with this kind of models (seesaw) is that have many free parameters, that can't be determinated by experiments, so it is necessary to create more ways to measure these parameters, or is necessary to theoretically reduce the model. We have chosen the second option, for this we have used an exchange symmetry between the $\mu$ and the $\tau$ flavor of the neutrinos. This symmetry is suggested due to the measured values in the atmospheric and the CHOOZ/PALOVERDE mixing angles, in the neutrino oscillations. Using this symmetry the parameters of the model are reduced so almost all can be determinated by experiments.

\vspace{.3cm}

Since we will have CP violation phases in the neutrino mass matrix, we have also propose the use leptogenesis to determine one more parameter. Leptogenesis is a model that explains the baryon-antibaryon asymmetry in the universe, we can use it's results to determine one phase in our model.

\vspace{.3cm}

Finally we did an approximation, for this we take values of the right handed neutrino masses in the form: $M_{3}>>M_{2}>>M_{1}$. With these we can make some predictions to compare with results from experiments, in concrete we give a value for the neutrinoless double beta decay.

\vspace{.3cm}

With the neutrinos theories born more than half a century ago, we still have more to learn about this elusive particle, and with it, learn more about the universe we live in; not only its present state, but also about it's beginning (the Big Bang), and it's future. Neutrinos may even help understanding dark matter.

\vspace{.3cm}

This thesis is just an academic exercise. Still, it is a good approximation to the understanding of this fascinating particle: ``The Neutrino''.
\chapter{Weak Interactions and the Neutrino}

The story of the neutrino\footnote{I actually don't mean a real neutrino, because many neutrinos exist since the Big Bang and many others are created all the time. What I mean is how the theories regarding neutrinos and many other particles were ``discovered''.} in physics starts many years before it was ``invented'', and eventually discovered; it starts with the discovery of the week interactions in 1896 by H. Becquerel \cite{Becquerel}; he found evidence for spontaneous radioactivity effect in uranium decay. One year later J.J. Thompson discovered the electron. This event gave birth to particle physics, it also gave a much better understanding of chemistry, electricity, and many other phenomena.

\vspace{.2cm}

The first decade of the 1900 was essential for the understanding of modern physics. In 1900 Planck started the quantum era and in 1905 Einstein started the relativistic era. The second decade gave birth to nuclear physics when Rutherford found evidence of an atomic nucleus in 1911. In that same year Millikan measured the electron charge. In 1913 Bohr invented the quantum theory for atomic spectra, and only one year later Chadwick \cite{Chadwick} found that the spectrum for the $\beta$ decay is continuous (which was confirmed in 1927 by Ellis and Wooster \cite{Ellis}). In 1919 Rutherford discovered that the atomic nucleus has positive charged particles, called protons.

\vspace{.2cm}

In the 1920's the theory of quantum mechanics was formulated. In 1923 de Broglie discovered the duality wave-corpse of the electron; in 1925 Pauli discovered the exclusion principle, and Heisenberg gave the structure of quantum mechanics; after that in 1926 Shr\"{o}dinger introduced the wave equation for quantum mechanics, and in 1927 and 1928 Dirac gave the foundations for Quantum Electrodynamics(QED)  \cite{Dirac}, and the relativistic wave equation that describes the electron.

\vspace{.2cm}

The forth decade was very important for the building of the weak interaction theory, also for nuclear and particle physics. In 1930 Pauli proposed the existence of a neutral particle emitted in $\beta$ decay \cite{Pauli1}; he called it neutron, but the name was changed four years later to neutrino, when he explained the $\beta$ decay \cite{Pauli2} as a neutron decaying into a proton, and electron and an anti-neutrino.
\begin{equation}
n\rightarrow p^{+} + e^{-} + \bar{\nu}.
\end{equation}

In 1931 Dirac predicted the anti-matter and only one year later evidence of the positron (the antiparticle of the electron) was found; in this same year (1932), Chadwick found evidence of the existence of the neutron (the reason for the change of name for the neutrino) which led to Heisenberg to postulate that the atomic nuclei is made of neutrons and protons. In 1934 Fermi postulated the Lagrangian that describes the $\beta$ decay, also referred as the field theory for $\beta$ decay \cite{Fermi}. In 1937 Majorana introduce his theory for neutrinos. In 1937 and in 1940 the muon and muon decay were observed \cite{Williams}.

\vspace{.2cm}

Between 1943 and 1949 the covariant QED theory was formulated by different people \cite{Tomonaga, Feynman, Schwinger, Tati, Dyson}. In 1949 it was proposed in two different articles \cite{Tiomno, Lee} that different processes of the same nature  must have the same coupling constant (Fermi constant). This is called the universality of Fermi weak interactions.

\vspace{.2cm}

At the end of the 1950's and after, there was a lot of work done perfecting the weak interaction theory, and making experiments that confirm it, by finding new particles, and measuring parameters.  In 1957 evidence for parity violation was found in some weak decays \cite{Wu, Garwin, Friedman}. But it wasn't until 1973 that the CP violation theory was integrated in the standard model \cite{Kobayashi}.

\vspace{.2cm}

The theory of different bosons as carriers of the weak interaction took about 30 years to be built, perfected and confirmed. It started in 1957, with just an idea \cite{Schwinger2, Yang}. In 1961 the neutral boson was proposed \cite{Glashow} and was discovered in 1973 \cite{Hasert}, which was a great success for the model. And in 1983 evidence of the charged boson was found~\cite{Arnison}.

In 1957 the neutrino theory was further developed, it was clear that the neutrino has to be either left or right handed \cite{Salam, Lee2, Landau}, one year later measurements of a negative helicity in the neutrino were done \cite{Goldhaber}, proving the left-handiness of the neutrino because the weak interaction had a V- A structure \cite{Feynman2, Marshak, Sakurai}. It is interesting to note that the neutrino wasn't discovered experimentally until many years after it was introduced; the electron neutrino was first detected in 1957, and the muon neutrino in 1962. And in 1958 and 1962 neutrino oscillations were proposed \cite{Theosci}.

\vspace{.2cm}

It was in the 60's and the beginning of the 70's when the Standard Model of particle physics was completed as we know it today. Goldstone and Higgs constructed a model for a spontaneous broken global symmetry of the Lagrangian, the first in 1961 for massive scalar bosons \cite{Goldstone}, and the second in 1964 for massive vector bosons \cite{Higgs}.

\vspace{.2cm}

In 1961 and 1964 Glashow, Gell-man, Salam and Ward \cite{Glashow, Glashow1, SalamWard, SalamWard2} used the gauge principle as a basis for quantum field theories of interacting fundamental fields and then proposed a Lagrangian for the unification of the weak and electromagnetic interactions. In that same year the quarks were ``discovered'' as fundamental particles that construct the hadrons (neutrons, protons, etc.). Finally in 1967 Weinberg gave the Lagrangian for electroweak unification \cite{Weinberg2},  and in 1973 Fritzsch, Gell-Mann and Leutwyler proposed the Lagrangian for Quantum Chromodynamics (QCD), that included the strong interaction in the model.

\vspace{.2cm}

It wasn't until many years later in 1989 that came the first experimental evidence on the number of light neutrinos \cite{Abrams}(for complete data of compiled experiments check \cite{lep}), and consequentially three lepton families. And in 1998 Super-Kamiokande Collaboration confirmed the existence of neutrino oscillations \cite{SuperK}.
\chapter{The Standard Model}

The object of this chapter is to make a description of the construction of the Standard Model (also known as the Glashow-Weinberg-Salam theory for the weak and electromagnetic interactions)\footnote{This chapter was primary written following \cite{cottin} and \cite{peskin}}. To understand the procedure some knowledge of quantum guage field theory is necessary as well as familiarity with the phenomenology of the weak interaction. For the propose of this thesis, it will be important to note the complete Lagangian for the interaction in section 3, and to realize that with out loss of generality this Lagrangian can be written in a diagonal base.

\section{The Glashow-Weinberg-Salam theory}

The matter sector of the Standard Model is composed of three families of quarks and leptons; they interact with each other through gauge bosons, which are spin-1 particles. These interactions are invariant under local symmetry transformations (this is the principle of gauge invariance).

The gauge group of the Standard Model is $ SU(3)_c \times SU(2)_L \times U(1)_Y$. Where the right handed leptons interact only with the $U(1)_Y$ bosons, and the left handed leptons interact under  $SU(2)_L \times U(1)_Y$.
$ SU(3)_c$ is the symmetry for the quark sector, where the bosons are the gluons.

For the propose of this thesis we are only going to make use of the leptons and the $SU(2)_L \times U(1)_Y$ symmetry, because we are not interested in the quark sector.

To begin with the building, first, the left $SU(2)$ doublet has to be defined, it consist on the left handed electron spinor $\ell_{L}$ and the left handed electron neutrino spinor $\nu_{L}$:
\begin{equation}
{\bf L}= \dbinom{\nu_{L}}{\ell_{L}} \label{dobleteiz}.
\end{equation}
\vspace{0.3cm}

This two particles seem to be very different, the electron has mass and a charge, and the neutrino has no mass in this model and has no charge. But it is important to note that due to the fact that a $SU(2)$ transformation mix the two spinor components of an $SU(2)$ doublet, only two spinor that have the same Lorentz transformation can be put together in a doublet, since the left handed electron and the left handed neutrino have the same transformation they are put in a doublet that has a transformation in the form:
\begin{equation}
{\bf L}\rightarrow {\bf L'}={\bf UL}\label{transdobleteiz}.
\end{equation}

Until today, a right handed neutrino has not been found, this have many consequences, including the fact that in this theory neutrinos doesn't have mass, as it will be explain in the next chapter. So the right handed electron spinor will be a singlet, in contrast with the left doublet. It can be concluded that the transformation for this singlet will be simply:
\begin{equation}
\ell_{R}\rightarrow \ell'_{R}=\ell_{R}\label{transsingleteder}.
\end{equation}
\vspace{-.2cm}

\section{Covariant Derivative}
We have introduce the leptons in the theory, now let's introduce a gauge field $W_{\mu}(x)$ this field is necessary to make the global $SU(2)$ symmetry into a local one. In the form of a 2 x 2 Hermitian matrix the field is:
\begin{equation} 
\textbf{W}_{\mu}(x)=W_{\mu}^k \sigma^k=
\left(
\begin{array}{cc}
W^{3}_{\mu} & (W^{1}_{\mu}-iW^{2}_{\mu})  \\
(W^{1}_{\mu}+iW^{2}_{\mu}) & -W^{3}_{\mu}  \end{array} 
\right)\label{bosonw},
\end{equation}
where $\sigma^{k}$ are the Pauli matrices.

Now we introduce a unitary matrix $U(x)$ that can be expressed in the following way:
\begin{equation} 
\textbf{U} = exp(-i\alpha^{k}\sigma^{k})\label{matrixuni},
\end{equation}
where $\alpha^{k}$ are three real numbers.

Using this matrix we can make the corresponding transformation for the $\textbf{W}_{\mu}(x)$:
\begin{equation} 
\textbf{W}_{\mu}(x)\rightarrow \textbf{W'}_{\mu}(x)= \textbf{U}(x)\textbf{W}_{\mu}(x)\textbf{U}^{\dagger}(x) + (2i)/g_{2})(\partial_{\mu} \textbf{U}(x))\textbf{U}^{\dagger}(x)\label{transformacionw}.
\end{equation}
The constant $g_{2}$ is a parameter of the model that can be determinated by phenomenology

In order to upgrade a global U(1) to a local one, it is necessary to introduce a gauge field $B_{\mu}$, it has the transformation:
\begin{equation}
B_{\mu} \rightarrow B'_{\mu} = B_{\mu} + 2/g_{1} \partial_{\mu} \theta \label{transfB}.
\end{equation}

With the help of this two fields we are able to write the covariant derivative for the left doublet and for the right singlet, this covariant derivative is introduce to maintain the $SU(2)\times U(1)$ symmetry. For the doublet:
\begin{equation}
D_{\mu} \textbf{L} = ( \partial_{\mu} + (ig_1 Y/2)B_{\mu} + (ig_{2}/2)\textbf{W}_{\mu})\textbf{L} \label{dercovdoub},
\end{equation}
and for the singlet:
\begin{equation}
D_{\mu} \ell_{R} = ( \partial_{\mu} + (ig_1 Y/2)B_{\mu})\ell_{R} \label{dercovsing},
\end{equation}
where $Y$ is the hypercharge, which is $-1$ for the left doublets and $-2$ for the right singlet. The field $\textbf{W}_{\mu}$ doesn't appear in the derivative for the singlet since it is the field associated with the $SU(2)$ transformation.

The next step in the Glashow-Weinberg-Salam theory is to define the Weak angle $\theta_{W}$; it is defined as a combination of the parameter constants $g_{2}$ and $g_{1}$ and can be determined with the phenomenology and the fact that we want to simplify the mass Lagrangian as it will be seen is the next section. This angle is defined by:
\begin{equation}
\cos\theta_{W} = \frac{g_{2}}{\sqrt{g_{1}^{2}+g_{2}^{2}}}\label{cosweinb},
\end{equation}
\begin{equation}
\sin\theta_{W} = \frac{g_{1}}{\sqrt{g_{1}^{2}+g_{2}^{2}}}\label{sinweinb}.
\end{equation}

The Weak angle is used to make a rotation in the axis, in the $(B_{\mu},W_{\mu}^{3})$ space, in the following way:
\begin{equation}
B_{\mu}=A_{\mu}\cos\theta_{W}-Z_{\mu}\sin\theta_{W}\label{rotB},
\end{equation}
\begin{equation}
W_{\mu}^{3}=A_{\mu}\sin\theta_{W}+Z_{\mu}\cos\theta_{W}\label{rotW}.
\end{equation}
This way we will be able to have a covariant derivative that will be a function of the physical fields $A_{\mu}$ and $Z_{\mu}$, where the first is related to the photon and the electromagnetic field, and the second is related to the Z boson, and the weak force.

\vspace{0.3cm}

It's also convenient to define $W_{\mu}^{+}$ and $W_{\mu}^{-}$ as a combination of $W_{\mu}^{1}$ and $W_{\mu}^{2}$ in the following way:
\begin{equation}
W_{\mu}^{+} = (W^{1}_{\mu}-iW^{2}_{\mu})/\sqrt{2}\label{Wplus},
\end{equation}
\begin{equation}
W_{\mu}^{-} = (W^{1}_{\mu}+iW^{2}_{\mu})/\sqrt{2}\label{Wminus}.
\end{equation}

Explicitly, the covariant derivative for the left handed doublet:
\begin{equation}
D_{\mu}\textbf{L}=\left[\partial_{\mu}\left(
\begin{array}{cc}
1 & 0  \\
0 & 1  \end{array} 
\right) + \frac{ig_{2}}{2}\left(
\begin{array}{cc}
W^{3}_{\mu} & \sqrt{2}W_{\mu}^{+}  \\
\sqrt{2}W_{\mu}^{-} & -W^{3}_{\mu}   \end{array} 
\right) + \frac{ig_{1}}{2}B_{\mu}\left(\begin{array}{cc}
1 & 0  \\
0 & 1  \end{array} 
\right)\right]
 \dbinom{\nu_{L}}{\ell_{L}}
\end{equation}
\[
=\left(\begin{array}{cc}
\partial_{\mu} + \frac{ig_{2}}{2}W^{3}_{\mu}- \frac{ig_1}{2}B_{\mu}& ig_{2}\sqrt{2}W_{\mu}^{+}  \\
ig_{2}\sqrt{2}W_{\mu}^{-} & \partial_{\mu} - \frac{ig_{2}}{2}W^{3}_{\mu}- \frac{ig_1}{2}B_{\mu}  \end{array} 
\right)\dbinom{\nu_{L}}{\ell_{L}}.
\]

Making the rotation of the axis it is easy to see that the covariant derivative will have the form:
\begin{equation}
D_{\mu}\textbf{L}= \left(\begin{array}{c}
\scriptstyle{\partial_{\mu}+ A_{\mu}(\frac{ig_{2}}{2}\sin\theta_{W}-\frac{ig_1}{2}\cos\theta_{W})+Z_{\mu}(\frac{ig_{2}}{2}\cos\theta_{W}+\frac{ig_1}{2}\sin\theta_{W})}  \hspace{.5cm} \scriptstyle{ig_{2}\sqrt{2}W_{\mu}^{+}}  \\
\scriptstyle{ig_{2}\sqrt{2}W_{\mu}^{-}} \hspace{.5cm}  \scriptstyle{\partial_{\mu}+ A_{\mu}(\frac{-ig_{2}}{2}\sin\theta_{W}-\frac{ig_1}{2}\cos\theta_{W})-Z_{\mu}(\frac{ig_{2}}{2}\cos\theta_{W}-\frac{ig_1}{2}\sin\theta_{W})}  \end{array} 
\right)\dbinom{\nu_{L}}{\ell_{L}}.
\end{equation}

The electron has charge -e, and the neutrino has neutral charge, so it is necessary that the term associated with the field $A_{\mu}$, will be only $-ieA_{\mu}$, for the electron, and zero for the neutrino. This impose values for the free parameters, it is necessary that:
\begin{equation}
g_1\cos\theta_{W}=g_{2}\sin\theta_{W}=-e \label{valueg}.
\end{equation}
Using these values the covariant derivative for the left doublet will finally be:
\begin{equation}
D_{\mu}\textbf{L}= \left( \begin{array}{cc}
\partial_{\mu} - \frac{ie}{\sin2\theta_{W}}Z_{\mu} & -\frac{ie}{\sqrt{2}\sin\theta_{W}}W_{\mu}^{+}  \\
-\frac{ie}{\sqrt{2}\sin\theta_{W}}W_{\mu}^{-}& \partial_{\mu}-ieA_{\mu}+(ie\cot2\theta_{W})Z_{\mu}  \end{array} 
\right)\dbinom{\nu_{L}}{\ell_{L}}\label{covdarleft}.
\end{equation}
In the same way, the covariant derivative for the right handed singlet will be: 
\begin{equation}
D_{\mu}\ell_{R}=(\partial_{\mu} - \frac{ig_1}{2}B_{\mu})\ell_{R},
\end{equation}
the derivative will be, after making the rotation (\ref{rotB}):
\begin{equation}
D_{\mu}\ell_{R}=\left[(\partial_{\mu}-ieA_{\mu})+ie\tan\theta_{W}Z_{\mu}\right]\ell_{R}\label{covdarright}.
\end{equation}
\vspace{-.1cm}

\section{Dynamical Part of the Lagrangian Density}
Using the covariant derivative we can write the dynamical part of the Lagrangian density  as a combination of the left and the right part
\begin{equation}
\mathscr{L}_{dyn}=\mathscr{L}_{dyn}^{L}+\mathscr{L}_{dyn}^{R}=\overline{\textbf{L}}i\gamma^{\mu} D_{\mu}\textbf{L}+ \overline{\ell_{R}}i\gamma^{\mu} D_{\mu}\ell_{R}.
\end{equation}

Making the corresponding matrix operations and including the covariant derivatives for the left handed doublet(\ref{covdarleft}), and the right handed singlet (\ref{covdarright}), the Lagrangian will finally be
\[
\mathscr{L}=\bar{\nu}_{L}i\gamma^{\mu}\partial_{\mu}\nu_{L}+\overline{\ell}_{L}i\gamma^{\mu}\partial_{\mu}\ell_{L}+ \overline{\ell}_{R}i\gamma^{\mu}\partial_{\mu}\ell_{R}\longrightarrow \mathscr{L}_{dyn}\]
\[+ e\bar{\ell}_{L}\gamma^{\mu}A_{\mu}\ell_{L}+ e\bar{\ell}_{R}\gamma^{\mu}A_{\mu}\ell_{R}\longrightarrow \mathscr{L}_{EM}\]
\[-\frac{e}{\sin2\theta_{W}}\bar{\nu}_{L}\gamma^{\mu}\nu_{L}Z_{\mu} + e\cot2\theta_{W}\overline{\ell}_{L}\gamma^{\mu}\ell_{L}Z_{\mu}-e\tan\theta_{W}\bar{\ell}_{R}\gamma^{\mu}\ell_{R}Z_{\mu}\longrightarrow \mathscr{L}_{Z}\]
\begin{equation}
-\frac{e}{\sqrt{2}\sin2\theta_{W}}\bar{\nu}_{L}\gamma^{\mu}\ell_{L}W_{\mu}^{+} - \frac{e}{\sqrt{2}\sin2\theta_{W}}\overline{\ell}_{L}\gamma^{\mu}\nu_{L}W_{\mu}^{-}\longrightarrow \mathscr{L}_{W}.
\end{equation}

An interesting fact about this Lagrangian is the term  $\mathscr{L}_{W}$, since it mixes the electron and the neutrino fields by the means of the W boson.

\vspace{0.3cm}

It is important to remember that there are three families of leptons. One can easily generalize the procedure done above by making the functions $\ell_{L}$, $\ell_{R}$ and $\nu_{L}$, into a $3 \times 1$ matrix vector, for the left handed charged lepton:
\begin{equation}
\ell_{L}=\left(\begin{array}{c}
e_{L}   \\
\mu_{L}   \\
\tau_{L}   \end{array} \right).
\end{equation}
For the right handed lepton:
\begin{equation}
\ell_{R}=\left(\begin{array}{c}
e_{R}   \\
\mu_{R}   \\
\tau_{R}   \end{array} \right).
\end{equation}
For the left handed charged neutrino:
\begin{equation}
\nu_{L}=\left(\begin{array}{c}
\nu_{eL}  \\
\nu_{\mu L}   \\
\nu_{\tau L}   \end{array} \right).
\end{equation}

We have formulated the Lagrangian for an $SU(2) \times U(1)$ theory, the next step is to add the mass term, that will complement the theory and give it physical meaning.

\section{The Higgs Mechanism}

The Lagrangian for a scalar field is:
\begin{equation}
\mathscr{L}_{scalar}=\partial_{\mu}\phi^{\dagger}\partial^{\mu}\phi- V(\phi^{\dagger}\phi)\label{scalarlag}.
\end{equation}
The potential has the form:
\begin{equation}
V(\phi^{\dagger}\phi)= \mu^2\phi^{\dagger}\phi+ \lambda(\phi^{\dagger}\phi)^2.
\end{equation}
The first term seems to be related to the mass, and the second to a self-interaction of the scalar field. The function $\phi$ is actually a doublet, and has the components:
\begin{equation}
\phi=\left(\begin{array}{c}
\phi^+ \\
\phi^0 \end{array}\right),
\end{equation}
with an expectation value in the vacuum (i.e, the minimum energy state):
\begin{equation}
\left\langle \phi\right\rangle=\left(\begin{array}{c}
0\\
v \end{array}\right),
\end{equation}
where $v$ can be calculated by the usual methods, using the energy as the function to minimize, leading to:
\begin{equation}
v=\sqrt{\frac{-\mu}{4\lambda}}.
\end{equation}

Now we will use the covariant derivative for a doublet (\ref{dercovdoub})
\begin{equation}
D_{\mu} \phi = ( \partial_{\mu} + (ig_1 /2)B_{\mu} + (ig_2 /2)W_{\mu}^k \sigma^k)\phi \label{derphi},
\end{equation}
where $Y=1$ for the Higgs doublet.

Looking at the Lagrangian (\ref{scalarlag}), we realize that to calculate the mass terms in the Lagrangian we have to evaluate the scalar field of the square of the derivative (\ref{derphi}), the mass terms will be:
\begin{equation}
\mathscr{L}_{m}=(0 \hspace{.3cm} v) ((-ig_{1} Y/2)B_{\mu} + (-ig_{2} /2)W_{\mu}^a \sigma^a) ( (ig_{1} /2)B^{\mu} + (ig_{2} /2)W^{b \mu} \sigma^b) \dbinom{0}{v}.
\end{equation}
making the product will lead to:
\begin{equation}
\mathscr{L}_{m}=\frac{v^2}{4}\left[ g_{2}^2(W_{\mu}^1)^2 + g_{2}^2(W_{\mu}^2)^2 + (-g_{2} W_{\mu}^3 + g_{1} B_{\mu})^2\right].
\end{equation}

Quadratic terms in a Lagrangian can be interpreted as mass terms, so there will be three massive vector bosons, two in the form:
\begin{equation}
W_{\mu}^{\pm}=\frac{1}{\sqrt{2}} (W_{\mu}^1 \mp W_{\mu}^2),
\end{equation}
which have a mass $m_W =gv/\sqrt{8}$, the third one will be:
\begin{equation}
Z_{\mu}=\frac{1}{\sqrt{g_{2}^2 + g_{1}^2}} (g_{2} W_{\mu}^3 - g_{1} B_{\mu}),
\end{equation}
which have a mass $m_z =\sqrt{g_{2}^2+g_{1}^2}v/\sqrt{8}$. Also there will be a massless vector field ($m_A =0$), which is the electromagnetic vector potential, and has the form:
\begin{equation}
A_{\mu}=\frac{1}{\sqrt{g_{2}^2 + g_{1}^2}} (g_{1} W_{\mu}^3 + g_{2} B_{\mu}).
\end{equation}

Now we will rewrite the covariant derivative for the fields we've just found, for this we will use $T^{k}$ as generators for the $SU(2)$ group instead of the Pauli matrices, so in terms of the mass eigentsate fields, the derivative will be, using $T^{\pm}=(T^1 \pm i T^2)$:
\begin{equation}
D_{\mu}=\partial_{\mu}-\frac{g_{2}}{\sqrt{2}}( W_{\mu}^{+} T^{+} + W_{\mu}^{-} T^{-})-i\frac{1}{\sqrt{g_{2}^2 + g_{1}^2}}Z_{\mu} (g_{2}^2 T^3 - g_{1}^2 Y)
\end{equation}
\[-i\frac{g_{2} g_{1}}{\sqrt{g_{2}^2 + g_{1}^2}}A_{\mu} (T^3 + Y).
\]

From the last term we can conclude that the massless gauge boson couples with the gauge generator $(T^3 + Y)$, which means that this gauge generator is the electric charge quantum number $Q$, we can also conclude that the electric charge $e$, must be:
\begin{equation}
e=\frac{g_{2} g_{1}}{\sqrt{g_{2}^2 + g_{1}^2}}\label{e}.
\end{equation}

Comparing equation (\ref{e}) with (\ref{valueg}) will lead us to conclude as mentioned in section 2 of this chapter that:
\begin{equation}
\cos\theta_{W} = \frac{g_{2}}{\sqrt{g_{1}^{2}+g_{2}^{2}}}\label{cosweinb2},
\end{equation}
\begin{equation}
\sin\theta_{W} = \frac{g_{1}}{\sqrt{g_{1}^{2}+g_{2}^{2}}}\label{sinweinb2}.
\end{equation}

Which is the definition for the Weak angle.

\chapter{Lepton masses}

Neutrino masses are not included in the Standard Model; the object of this chapter is to introduce the different types of mass terms that can be added to the Lagrangian if a right handed neutrino is introduced to the model, this will mean new physics. At the end of the chapter we want to have built a mass Lagrangian for all the leptons. We will also observe that there are Dirac and Majorana particles, each with different mass terms. Finally we will introduce the seesaw model that will predict light masses for the left handed neutrinos if the right handed neutrinos  have heavy masses.

\vspace{-.3cm}

\section{Dirac Vs Majorana}
The electron, the muon and the tau are the charged leptons, these are called Dirac particles (though they are not the only ones), because they obey the Dirac equation. This kind of particles can be described by a four component spinor; two components for the left and right chiriality particles, and two for the right and left handed antiparticles.

\vspace{0.3cm}

In the case of neutrinos there are three possibilities, the first one is that if neutrinos are massless they can be described with a two component spinor (Weyl spinor), but we know neutrinos must have mass (depending on the hierarchy of the masses one can be zero), so this is not an option. The second is that neutrinos have the same behavior as the charged leptons, so they must be Dirac particles described by Dirac spinors. But there is a third option, if neutrinos are their own antiparticles (which can be possible since neutrinos don't have charge), then a two component spinor (Majorana spinor) can be use to describe them.

\vspace{0.3cm}

Now let's explain the chiriality projector operators (for reviews read: \cite{kayser2, Bilenky, Bilenky2}); these are projectors that make the complete wave function to transform into an either left or right wave function, and if the wave function already has a chiriality, then it stays the same or is zero depending if the projector is of the same chiriality of the opposite one, this means:
\begin{equation}
P_{R,L}\psi= \psi_{R,L}   ,  
\end{equation}
\[P_{R,L}\psi_{R,L}=\psi_{R,L}, \] 
\[P_{L}\psi_{L}=\psi_{L}, \] 
\[P_{L,R}\psi_{R,L}=0, \] 
\[P_{R,L}\psi_{R,L}=\psi_{R,L}, \] 
where $P_{R,L}=(1\pm \gamma_{5})/ 2$ is the chiriality operator.

\vspace{.3cm}

Mass terms in the Lagrangian may convert terms from one chiriality to another, or even to a different flavor, in the case of neutrinos there are two different kinds of terms for the mass, the Dirac term and the Majorana term. For this we will consider the wave functions $\nu_{L}$ and  $\nu_{R}$, but also the charged conjugated wave functions $(\nu_{L})^{c}$  and $(\nu_{R})^{c}$, defined as:
\begin{equation}
(\nu_{L})^{c}\equiv C\overline{\nu_{L}}^{T} ,\hspace{.2 cm} (\nu_{R})^{c}\equiv C\overline{\nu_{R}}^{T} \label{conjwave}.
\end{equation}
Here, C is the charge conjugation operator and has the following properties:
\begin{equation}
CC^{\dagger}=1, \hspace{.2 cm} C^{T}=-C, \hspace{.2 cm} C\gamma_{\alpha}^{T}C^{-1}=-\gamma_{\alpha}. \label{conjoper}
\end{equation}
And consequentially form (\ref{conjwave}) and (\ref{conjoper}) :
\begin{equation}
\overline{(\nu_{L})^{c}}=-\nu_{L}^{T}C^{-1}, \hspace{.3 cm} \overline{(\nu_{R})^{c}}=-\nu_{R}^{T}C^{-1}.
\end{equation}

Now let's use these wave functions to produce the mass term in the Lagrangian, the first one is the Dirac mass term, and it's constructed with the non-charged conjugate terms $\nu_{L}$ and  $\nu_{R}$
\begin{equation}
\mathscr{L}^{D}=-\overline{\nu_{R}}M_{D}\nu_{L}+ H.c.\label{diracmass}
\end{equation}

The second term is the Majorana mass term, and is constructed between a charged conjugated wave function and a non-charged conjugated.
\begin{equation}
\mathscr{L}^{M}=-\frac{1}{2}\overline{(\nu_{R})^{c}}M_{M}\nu_{R}+ H.c.\label{majorana mass}
\end{equation}
Another possible mass term can be:
\begin{equation}
\mathscr{L}^{ML}=-\frac{1}{2}\overline{(\nu_{L})^{c}}M_{L}\nu_{L}+ H.c. \label{massL}
\end{equation}
All other possible combinations can be reduced to one of these three terms, for example:
\begin{equation}
\overline{(\nu_{L})^{c}}M^{D}(\nu_{R})^{c}=-\nu_{L}^{T}C^{-1}M^{D}C \overline{\nu_{R}}^{T}= \overline{\nu_{R}}(M^{D})^{T}\nu_{L}.
\end{equation}

\section{Yukawa Couplings}

Gauge invariance prevents the existence of mass terms in the Lagrangian for standard leptons. In order to generate mass we introduce Yukawa interaction terms \cite{mohabook2}, that are allowed by gauge invariance, and then use the vacuum expectation values, to break the symmetry and generate the mass. 

The Yukawa Lagrangian for the $SU(2) \times U(1)$ model will be (see for instance \cite{gonzalez, cheng}):

\begin{equation}
-\mathscr{L}_{Y}=\sum_{a,b}\left[ h_{ab}^{(\ell)}\overline{\textbf{L}_{a}}\phi \ell_{bR}\right] + H.c.
\end{equation}
Where $\phi$ is the Higgs doublet, $\textbf{L}$ is the left lepton doublet, $\ell_{R}$ is the right lepton singlet, $h^{(\ell)}$ is the coupling matrix, the indices a and b are lepton family indices $\phi$ is the Higgs doublet. 

Breaking the symmetry by introducing the vacuum expectation value will led to the mass Lagrangian:
\begin{equation}
-\mathscr{L}_{mass}=\sum_{a,b}\left[ \overline{\ell_{aL}}M_{ab}^{\ell}\ell_{bR}\right] + H.c.
\end{equation}
Where the mass term for the charged leptons is defined as:
\begin{equation}
M_{ab}^{(\ell)}=h_{ab}^{(\ell)}v.
\end{equation}

In the same manner we can introduce a Yukawa coupling to generate Dirac mass terms for neutrinos
\begin{equation}
-\mathscr{L}_{Y}=\sum_{a,b}\left[h_{ab}^{(\nu)}\overline{\textbf{L}_{a}} \hat{\phi}\nu_{bR} \right],
\end{equation}
where:
\begin{equation}
\hat{\phi}=i\sigma_{2}\phi^{*}.
\end{equation}
Which after symmetry breaking will lead to the Dirac mass term (\ref{diracmass}).

An important propriety of the Majorana mass matrix it is that it is symmetric with respect to its diagonal, which means that in general it has 12 independent parameters (6 real masses, and 6 phases). The demonstration for this can been seen in reference~\cite{mohabook4}.

\section{Complete Mass Lagrangian Density}

Using the mass terms introduced previously we can construct the mass Lagrangian for the leptons, it will be:
\begin{equation}
\mathscr{L}_{mass}= \mathscr{L}^{\ell} + \mathscr{L}^{D} + \mathscr{L}^{M} =\sum_{a,b}\left[-\overline{\ell_{aL}}M_{ab}^{\ell}\ell_{bR} -\overline{\nu_{aL}}M^{D}_{ab}\nu_{bR} -\frac{1}{2}\overline{(\nu_{aR})^{c}}M^{M}_{ab}\nu_{bR}\right]+ H.c.
\end{equation}
The left Majorana term (\ref{massL}) has not been used, the reason to do this is due to a practical approach: We are looking for simplicity. This term cannot be generated by the couplings to standard Higgs at the renormalizable level. Not using this term will make a simpler model, will prevent adding new particles and most important will allow us to use the seesaw mechanism which will lead to some interesting predictions, as it will be seen in next section.

\vspace{0.3cm}

Taking only the neutrino mass terms in the Lagrangian, and introducing the new base
\begin{equation}
N=\left(\begin{array}{c}
(\nu_{L})^{c} \\
\nu_{R} \end{array}\right).
\end{equation}
Then we can write the Lagrangian as:
\begin{equation}
-\mathscr{L}_{mass}=\overline{\ell_{L}}M_{\ell} \ell_{R} + \frac{1}{2}\overline{N^{c}}MN +H.c ,
\end{equation}
where M is a 2 x 2 matrix in the form:
\begin{equation}
M=\left( \begin{array}{cc}
0 & M_{D} \\
M_{D}^{T} & M_{M} \end{array}\right),
\end{equation}
where $M_{D}$ and $M_{M}$ are in general 3 x 3 non-diagonal matrices.

\section{Seesaw Mechanism}

It's easy to diagonalize the mass matrix M. For this we will need to calculate he eigenvalues, and for simplicity we have chosen to use the matrices $M_{M}$ and $M_{D}$ to be 1 x 1, and then generalize to the case 3 x 3, the secular equation will be:
\begin{equation}
m_{1,2}= \frac{M_{M}\pm \sqrt{(M_{M})^{2}+\frac{4M_{D}^{2}}{M_{M}^{2}}}}{2}.
\end{equation}

Now we can make an approximation, for this we will make the assumption that $M_{M} >> M_{D}$, that can be made just by arguing that nature makes this hierarchy and remembering that if  $x<<1$ then:
\[\sqrt{1+x}\approx 1 +x/2.\]
This will lead to two different masses
\begin{equation}
m_{1}\approx M_{M} + \frac{M_{D}^{2}}{M_{M}}\approx M_{M}\equiv M_{R},
\end{equation}
and
\begin{equation}
m_{2}\approx - \frac{M_{D}^{2}}{M_{M}}\equiv m_{L}.
\end{equation}

This is called the seesaw mechanism \cite{seesaw}, and what it actually means is that since $M_{M} >> M_{D}$, by this mechanism we have generated two masses, where $m_{1}$ is much much bigger than $m_{2}$, so there will be two kinds of neutrinos, the left handed neutrino which is the lightest particle (below the quark and electron mass) and the right handed neutrino which is a very heavy particle (above the Higgs boson mass).

\vspace{0.3cm}

The mass Lagrangian for the left and right handed neutrinos in this new base \footnote {We maintain the same notation for the right and left handed neutrinos because even tough we have a new base, the mixture will be very small, that will be on the order of $M_D / M_M$ $\sim O(10^{-11})$}, for a single generation, will simply be:
\begin{equation}
-\mathscr{L}_{mass}=m_{L}\overline{({\nu}_{L})^c} \nu_{L} + M_{R}\overline{({\nu}_{R})^c} \nu_{R} + H.c.
\end{equation}

A simple calculation can demonstrate that in this base, both particles are Majorana particles \cite{mohabook3} and that for 3 generations we will have 6 particles, where the seesaw mass matrix for the left handed fields, will be (at a the lower order):
\begin{equation}
M_{\nu}=M_{D}\frac{1}{M_{M}}M_{D}^{T}\label{seemass1}.
\end{equation}

\chapter{Neutrino oscillations}

Neutrino oscillations have been confirmed \cite{osci}, but the theory regarding neutrino oscillations was proposed almost fifty years ago, by Pontecorvo, Maki, Nakagawa and Sakata, giving the mixing matrix the name PMNS \cite{Theosci}.
The objective of this chapter is to observe some of the proprieties of this oscillations. Also we will count the number of parameters that the mixing matrix must have  so we can construct it explicitly. Finally we will observe how this matrix is used in the Lagrangian for the theory. 

\section{Mass and Flavor Bases}

To start the theory of neutrino oscillations we have to assume that there are two different bases for neutrino wave functions, the first is for the mass eigenstates, in this base neutrinos have a definite mass and are freely propagating; the other base is for a combination of the mass eigenstates and are the electron ($\nu_{e}$), muon ($\nu_{\mu}$) and tau ($\nu_{\tau}$) neutrinos.
This can be written  in terms of each other in the following way:
\begin{equation}
\nu_{\ell}= \sum^{3}_{m=1} U_{\ell m}\nu_{m}\label{superposition},
\end{equation}
where the index $\ell$ can be e, $\mu$ or $\tau$, and the index $m$ represents the three mass eigenstates, and U is called the mixing matrix.

\section{Neutrino oscillations in Vacuum}

The wave function for a neutrino with defined momentum $p$ at a time $t=0$ is \cite{kayser}
\begin{equation}
\psi(x)= \nu_{\ell} e^{ipx}.
\end{equation}
Using the superposition (\ref{superposition}):
\begin{equation}
\psi(x)= \sum_{m}U_{\ell m}\nu_{m} e^{ipx},
\end{equation}
and the relativistic energy:
\begin{equation}
E_{m}=\sqrt{p_{\nu}^{2}+m_{m}^{2}},
\end{equation}
the wave function $\psi$ will be after a time t
\begin{equation}
\psi(x,t)=\sum_{m}U_{\ell m}\nu_{m} e^{ipx} e^{-iE_{m}t}.
\end{equation}

Since neutrinos travel at near speed of light we can make the approximation $t \approx x$, so:
\begin{equation}
\psi(x,x)\approx \sum_{m}U_{\ell m}\nu_{m} e^{-i\left[m_{m}^{2}/2p_{\nu}\right]x} ,
\end{equation}
where we have used the approximation $m_{m} << p_{\nu}$, and consequently:
\begin{equation}
E_{m} \approx p_{\nu} + \frac{m_{m}^{2}}{2p_{\nu}}.
\end{equation} 
This approximation can be done because neutrinos have very small masses, but high momentum due to very high speed. 

\vspace{0.3cm}

If we express $\nu_{m}$ as a linear combination of $\nu_{\ell}$ we get:
\begin{equation}
\psi(x,x)= \sum_{\ell'} \left[\sum_{m} U_{\ell m} e^{-i(m_{m}^{2}/2p_{\nu})x}U^{*}_{\ell'm}\right]\nu_{\ell'}\label{psi},
\end{equation}
where:
\begin{equation}
\nu_{m}= \sum_{\ell'} U^{*}_{\ell' m}\nu_{\ell'}.
\end{equation}
One can see in (\ref{psi}) that the coefficient has a relation with the probability of a neutrino $\nu_{\ell}$ to have a new flavor $\ell'$ after traveling a distance x, this will be

\begin{equation}
P(\ell\rightarrow \ell',x) = \left[\sum_{m'} U^{*}_{\ell m'} e^{i(m_{m'}^{2}/2p_{\nu})x}U_{\ell'm'}\right]\cdot \left[\sum_{m} U_{\ell m} e^{-i(m_{m}^{2}/2p_{\nu})x}U^{*}_{\ell'm}\right]
\end{equation}

\[=\sum_{m}\left|U_{\ell m}\right|^{2}\left|U_{\ell' m}\right|^{2}
+ \sum_{m'\neq m} Re(U_{\ell m}U^{*}_{\ell m'}U_{\ell' m'}U^{*}_{\ell' m})\cos(\frac{m^{2}_{m}-m^{2}_{m'}}{2p_{\nu}}x)\]

\[+ \sum_{m'\neq m} Im(U_{\ell m}U^{*}_{\ell m'}U_{\ell' m'}U^{*}_{\ell' m})\sin(\frac{m^{2}_{m}-m^{2}_{m'}}{2p_{\nu}}x).\]

We can see that the probability of a neutrino oscillating between flavors has to do with the difference in the squared masses $\Delta m^{2}$, and the distance the neutrino travels. So experimentally it is clear that we can't measure directly the mass of a neutrino in neutrino oscillations experiments \footnote{We could measure directly the mass of neutrino in the beta decay, by measuring the energy difference between the neutron with the proton and the electron (leaving only the energy of the neutrino), but this require extremely sensitive detector and the technology doesn't exists yet.}, but we can measure $\Delta m^{2}$.

\section{Mixing matrix}

To determine the mixing matrix $U$ of (\ref{superposition}) let's begin by counting the number of independent parameters the matrix can be expressed with \cite{chaichian}.

\vspace{0.3cm}

First, let's start by remembering that we have three families of lepton doublets, this means there are six leptons: electron, muon, tau, electron neutrino, muon neutrino and tau neutrino. This will make the mixing matrix a $3 \times 3$ matrix that in principle will have 18 parameters: nine real parameters and nine imaginary phases. $U$ must be unitary ( i.e $UU^{\dagger}=1$), this impose nine constrains in the matrix. This will make our count for independent parameters to be nine. 

\vspace{0.3cm}

For every wave function there is an unphysical phase that can be removed, since there are six wave functions, then in principle we could remove six more parameters, but one of the phases can be chosen to be zero, so then we will have five parameter to remove, and that will leave $9 - 5= 4$ independent parameters.

\vspace{0.3cm}

Let's see this explicitly; from the four parameters, three are angles and one is a phase; the angles are used to represent a rotation between a neutrino of one flavor to another flavor. We choose the parameterization used in \cite{chau}, so the mixing matrix will be:
\[
U=\left(
\begin{array}{ccc}
1 & 0 & 0  \\
0 & c_{y} & s_{y}  \\
0 & -s_{y} & c_{y}   \end{array} 
\right)\cdot \left(
\begin{array}{ccc}
c_{z} & 0 & s_{z} e^{i\delta} \\
0 & 1 & 0  \\
-s_{z} e^{-i\delta} & 0 & c_{z}   \end{array} 
\right)\cdot \left(
\begin{array}{ccc}
c_{x} & s_{x} & 0  \\
-s_{x} & c_{x} & 0  \\
0 & 0 & 1   \end{array} 
\right).
\]

\begin{equation}
=\left(\begin{array}{ccc}
c_{x}c_{z} & s_{x}c_{z} & s_{z} e^{-i\delta}  \\
-s_{x}c_{y}-c_{x}s_{y}s_{z} e^{i\delta} & c_{x}c_{y}-s_{x}s_{y}s_{z} e^{i\delta} & s_{y}c_{z}  \\
s_{x}s_{y}-c_{x}c_{y}s_{z} e^{i\delta} & -c_{x}s_{y}-s_{x}c_{y}s_{z}  e^{i\delta}& c_{y}c_{z}  \end{array} 
\right).
\end{equation}

Making a simple change of notation: $c_x=c_{12}, c_y=c_{23}, c_z=c_{13}, s_x=c_{12}, s_y=c_{23}$ and $s_z=c_{13}$ (This change is made because is more easy to understand the oscillations as a change form neutrino i to a neutrino j, and not as a rotation of axis). And with an adequate change in the phase for the angles, we get \cite{mohabook}:
\begin{equation}
U=\left(
\begin{array}{ccc}
c_{12}c_{13} & -s_{12}c_{13} & c_{12}s_{13} e^{-i\varphi} \\
s_{12}c_{23} + c_{12}s_{23}s_{13} e^{i\varphi} & c_{12}c_{23} - s_{12}s_{23}s_{13} e^{i\varphi} & -s_{23}c_{13} \\
s_{12}s_{23} - c_{12}c_{23}s_{13} e^{i\varphi} & c_{12}s_{23} + s_{12}c_{23}s_{13} e^{i\varphi} & c_{23}c_{13} \end{array} 
\right)\label{mns}.
\end{equation}

Due to the $U(1)$ symmetry of the wave function for leptons, we could multiply the matrix $U$ by a diagonal matrix of phases: 
\begin{equation}
K=\left(
\begin{array}{ccc}
e^{i\alpha} & 0 & 0 \\
0 & e^{i\beta}  & 0 \\
0 & 0 & e^{i\gamma} \end{array} 
\right),
\end{equation}
so finally:
\begin{equation}
U_{PMNS}=U \cdot K \label{UPMNS}.
\end{equation}

The matrix of phases $K$ is introduced because we want the Majorana mass matrix to have real eigenvalues, so by introducing this matrix we will move the phases form the Majorana mass matrix to the mixing matrix. But due to a global $U(1)$ symmetry the diagonal Majorana mass matrix only has two physical complex phases, so one of the phases is set to zero, and this will leave only three independent phases (one Dirac phase, and two Majorana phases). But one can choose to leave the phases in the Majorana mass matrix if it helps to reduce parameters in future calculations. 

\vspace{0.3cm}

Now we will show the exact part of the theory were the mixing matrix is used, for this we will start with the complete Lagrangian for the lepton model
\begin{equation}
\mathscr{L}=\bar{\nu}_{L}i\gamma^{\mu}\partial_{\mu}\nu_{L}+\bar{\ell}_{L}i\gamma^{\mu}\partial_{\mu}\ell_{L}+\bar{\ell}_{R}i\gamma^{\mu}\partial_{\mu}\ell_{R} + e\bar{\ell}_{L}\gamma^{\mu}A_{\mu}\ell_{L}+ e\bar{\ell}_{R}\gamma^{\mu}A_{\mu}\ell_{R}
\end{equation}
\[ -\frac{e}{\sin2\theta_{W}}\bar{\nu}_{L}\gamma^{\mu}\nu_{L}Z_{\mu} + e\cot2\theta_{W}\overline{\ell}_{L}\gamma^{\mu}\ell_{L}Z_{\mu}- e\tan\theta_{W}\bar{\ell}_{R}\gamma^{\mu}\ell_{R}Z_{\mu} \]
\[ -\frac{e}{\sqrt{2}\sin2\theta_{W}}\bar{\ell}_{L}\gamma^{\mu}\nu_{L}W_{\mu}^{-} -\overline{\ell}_{L}M_{\ell}\ell_{R} -\overline{\nu_{L}}M_{D}\nu_{R} -\frac{1}{2}\overline{(\nu_{R})^{c}}M_{M}\nu_{R} + H.c.
\]

And now we will try to diagonalize as much as we can. To be in a diagonal base is important because a non-diagonal base will mean that there is some kind of interaction or oscillation between the 3 families, and for example we know that the dynamical part of the Lagrangian must not have this kind of interactions, also for example we know to a very exact limit, the masses for the three charged leptons, so the part of the Lagrangian for these masses will have to be in a mass eigenstate base, which means that the mass matrix for these should be diagonal with three different eigenstates. 

\vspace{0.3cm}

With out loss of generality, it can be assumed that all the terms in the Lagrangian are already on a diagonal base, but the three mass terms. To try to diagonalize, as much as we can, we will only take four parts of the Lagrangian, the interaction with the W boson (which is already in a diagonal base) and the mass terms, so we will have the following Lagrangian: 
\begin{equation}
-\mathscr{L}=  \overline{\ell}_{L}M_{\ell}\ell_{R} +\overline{\nu_{R}}M_{D}\nu_{L} +\frac{1}{2}\overline{(\nu_{R})^{c}}M_{M}\nu_{R}+\frac{e}{\sqrt{2}\sin2\theta_{W}}\overline{\nu}_{L}\gamma^{\mu}\ell_{L}W_{\mu}^{+}+H.c.
\end{equation}

Now we can introduce some unitary matrices (or orthogonal) by pairs in strategic places, so the Lagrangian would be:
\begin{equation}
\mathscr{L}=  -\overline{\ell}_{L}U_{L}U_{L}^{\dagger}M^{\ell}U_{R}U_{R}^{\dagger}\ell_{R} -\overline{\nu_{R}}V_{R}V_{R}^{\dagger}M^{D}V_{L}V_{L}^{\dagger}\nu_{L} -\frac{1}{2}\overline{(\nu_{R})^{c}}V_{R}V_{R}^{T}M^{M}V_{R}V_{R}^{T}\nu_{R}
\end{equation}
\[-\frac{e}{\sqrt{2}\sin2\theta_{W}}\overline{\nu}_{L}\gamma^{\mu}V_{L}U_{L}^{\dagger}\ell_{L}W_{\mu}^{+}+H.c.\]
and make the following transformations for the masses:
\begin{equation}
M_{\ell}^{Diag}=U_{L}^{\dagger}M_{\ell}U_{R}, \hspace{.3cm}
M_{D}^{Diag}=V_{R}^{\dagger}M_{D}V_{R}, \hspace{.3cm}
M_{M}^{Diag}=V_{R}^{T}M_{M}V_{R} ,
\end{equation}
and for the wave functions:
\begin{equation}
\overline{\ell}_{L}U_{L}=\overline{\ell}'_{L}, \hspace{.3cm}
\overline{\ell}_{R}U_{R}=\overline{\ell}'_{R}, \hspace{.3cm}
\overline{\nu_{R}}V_{R}=\overline{\nu_{R}}', \hspace{.3cm}
\overline{\nu_{L}}V_{L}=\overline{\nu_{L}}'.
\end{equation}

The only part of the Lagrangian that was not diagonalized was the interaction with the $W$ boson, and the mixing matrix will be:

\begin{equation}
V_{L}U_{L}^{\dagger}=U_{PMNS}.
\end{equation}

This is the PMNS Matrix (\ref{UPMNS}) that has been introduced in this chapter.
\chapter{$\mu$-$\tau$ symmetry}

In this chapter we will explain why we are interested in this symmetry, first we will make a unitary transformation on a diagonal matrix using the mixing matrix and experimentally measured values of two of the angles, this will suggest that the mass matrix for neutrino might have a $\mu$-$\tau$ symmetry. Then we will construct the Majorana and the Dirac mass matrices using this symmetry. We will find that one of the angles is related to a combination of the components of the Majorana mass matrix, and finally we will find equations that relate the mass of the neutrinos in the mass eigenstate and the components of the Majorana mass matrix in the flavor base.

\section{Why $\mu$-$\tau$ symmetry?}

The PMNS matrix (\ref{mns}) is:
\begin{equation}
U_{PMNS}=\left(
\begin{array}{ccc}
c_{12}c_{13} & -s_{12}c_{13} & c_{12}s_{13} e^{-i\delta} \\
s_{12}c_{23} + c_{12}s_{23}s_{13} e^{i\delta} & c_{12}c_{23} - s_{12}s_{23}s_{13} e^{i\delta} & -s_{23}c_{13} \\
s_{12}s_{23} - c_{12}c_{23}s_{13} e^{i\delta} & c_{12}s_{23} + s_{12}c_{23}s_{13} e^{i\delta} & c_{23}c_{13} \end{array} 
\right).
\end{equation}
where $c_{12}\equiv\cos\theta_{12}$, $c_{13}\equiv\cos\theta_{13}$, $c_{23}\equiv\cos\theta_{23}$ and $s_{12}\equiv\sin\theta_{12}$, $s_{13}\equiv\sin\theta_{13}$, $s_{23}\equiv\sin\theta_{23}$.

The Super-Kamiokande experiment \cite{SuperK} gave for the first time in 1998 irrefutable evidence of neutrino oscillation, current data obtained from different experiments (i.e CHOOZ, PALOVERDE, MINOS and KamLand) suggest that the $\mu$ and $\tau$ neutrino are maximally mixed in atmospheric neutrino oscillations \cite{bimaximal2, bimaximal3, bimaximal}, and that there is null mixing between electron neutrinos and tau neutrinos in atmospheric measurements, the combined data of all the experiments \cite{maltoni} gives the following results:
\begin{equation}
\theta_{12}=33.7\pm 1.3, \hspace{.3cm} \theta_{ATM}=\theta_{23}=43.3_{-3.8}^{+4.3}, \hspace{.3cm} \theta_{13}=0_{-0.0}^{+5.2}.
\end{equation}

These results give the idea that we could use the following angles:
\begin{equation}
\theta_{23}\approx\pi/4, \hspace{.4cm}\theta_{13}\approx 0.
\end{equation}
\vspace{.2cm}
With these values, the mixing matrix (with out the complex Majorana phases) is rewritten as:
\begin{equation}
U_{PMNS}^{(\mu \leftrightarrow \tau)}=\left(
\begin{array}{ccc}
\cos\theta_{12} & -\sin\theta_{12} & 0 \\
\sin\theta_{12}/\sqrt{2} & \cos\theta_{12}/\sqrt{2} & -1/\sqrt{2} \\
\sin\theta_{12}/\sqrt{2} & \cos\theta_{12}/\sqrt{2} & 1/\sqrt{2} \end{array} \right)\label{PMNS}.
\end{equation}

The $\delta$ phase (called Dirac phase) have disappeared from the mixing matrix, this is due to the fact that this phase is always multiplied by the $\sin \theta_{13}$, in the limit where this angle is zero, then the sine goes to zero as well. But we have to be careful because this happens only in the limit, if $\theta_{13}$ is slightly different form zero, then the phase will not disappear, and will be another parameter for the model.

\vspace{0.3cm}

Matrix (\ref{PMNS}) diagonalizes the mass matrix, so if we have a non diagonal matrix M we can transform it into a diagonal one by the operation:
\begin{equation}
M^{diag}= U^{\dagger}MU \label{diagmatrix}.
\end{equation}
Which implies the inverse operation:
\begin{equation}
M= UM^{diag}U^{\dagger}\label{mnondiag}.
\end{equation}
Where $M^{diag}$ a diagonal non-degenerate mass matrix
\begin{equation}
M^{diag}=\left( \begin{array}{ccc}
M_1 & 0 & 0 \\
0 & M_2 & 0 \\
0 & 0 & M_3 \end{array}\right).
\end{equation}
And using the PMNS matrix, the non-diagonal mass matrix will be, after applying operation (\ref{mnondiag}):
\begin{equation}
M=\left(\begin{array}{ccc}
\scriptscriptstyle{M_1\cos^{2}\theta_{12} + M2\sin^{2}\theta_{12}} & \scriptscriptstyle{(M_1-M_2)\cos\theta_{12}\sin\theta_{12}/\sqrt{2}} & \scriptscriptstyle{(M_1-M_2)\cos\theta_{12}\sin\theta_{12}/\sqrt{2}} \\
\scriptscriptstyle{(M_1-M_2)\cos\theta_{12}\sin\theta_{12}/\sqrt{2}} & \scriptscriptstyle{(M_3+M_2)\cos^{2}\theta_{12}+M1\sin^{2}\theta_{12}/2} & \scriptscriptstyle{(-M_3+M_2)\cos^{2}\theta_{12}+M1\sin^{2}\theta_{12}/2} \\
\scriptscriptstyle{(M1-M_2)\cos\theta_{12}\sin\theta_{12}/\sqrt{2}} & \scriptscriptstyle{(-M_3+M_2)\cos^{2}\theta_{12}+M1\sin^{2}\theta_{12}/2 }& \scriptscriptstyle{(M_3+M_2)\cos^{2}\theta_{12}+M1\sin^{2}\theta_{12}/2} \end{array}\right).\label{massmuitau}
\end{equation}
Which has the general form:
\begin{equation}
M=\left( \begin{array}{ccc}
a & b & b \\
b & c & d \\
b & d & c \end{array}\right).\label{mass}
\end{equation}
This matrix clearly has a $\mu-\tau$ symmetry because the exchange of the elements $M_{e \mu}$ with $M_{e \tau}$, $M_{\mu e}$ with $M_{\tau e}$, $M_{\mu \mu}$ with $M_{\tau \tau}$ and $M_{\mu \tau}$ with $M_{\tau \mu}$ leaves the matrix intact. This gives the idea that there is a hidden $\mu-\tau$ symmetry in the neutrino sector, that has to be broken, because the symmetry does not exists in the charged lepton sector. \textit{ i.e.} the muon mass is different to the tau mass.

\section{Majorana and Dirac Mass Matrices}

The general form for mass matrix related to the neutrinos is:
\begin{equation}
M= \left(\begin{array}{ccc}
M_{ee} & M_{e \mu} & M_{e \tau} \\
M_{\mu e} &  M_{\mu \mu} & M_{\mu \tau} \\
M_{\tau e}  &  M_{\tau \mu} & M_{\tau \tau} \end{array}\right).
\end{equation}
Then the explicit form of the mass matrices, when there is a $\mu- \tau$ symmetry involved, can be calculated noting that there are going to be equal terms in the matrix, this are: $M_{e \mu}=M_{e \tau}$, $M_{\mu \mu}=M_{\tau \tau}$, $M_{\mu e}=M_{\tau e}$ and $M_{\mu \tau}=M_{\tau \mu}$.  Using these we will have that the Majorana mass matrix has the form:
\begin{equation}
M_{M}^{(\mu \leftrightarrow \tau)}= \left(\begin{array}{ccc}
M_{ee} & M_{e \mu} & M_{e \mu} \\
M_{e \mu} &  M_{\mu \mu} & M_{\mu \tau} \\
M_{e \mu}  &  M_{\mu \tau} & M_{\mu \mu} \end{array}\right),\label{mmaj}
\end{equation}
whereas the Dirac mass matrix becomes:
\begin{equation}
M_{D}^{(\mu \leftrightarrow \tau)}= \left(\begin{array}{ccc}
D_{ee} & D_{e \mu} & D_{e \mu} \\
D_{\mu e} &  D_{\mu \mu} & D_{\mu \tau} \\
D_{\mu e}  &  D_{\mu \tau} & D_{\mu \mu} \end{array}\right).\label{mdirac}
\end{equation}

The only difference between the form of this two matrices is that the Majorana matrix is symmetrical with respect to the diagonal (as explained in chapter 3), and the Dirac matrix isn't, so for the Majorana mass matrix the condition $M_{\mu e}=M_{\tau e}$ was actually not used. This means that the Majorana matrix will have in principle 8 independent parameters (4 real masses and 4 complex phases) and the Dirac matrix will have 10 (5 real masses and 5 complex phases). The $\mu$-$\tau$ symmetry had reduced the number of parameters for both the matrices.

\section{Value for $\sin\theta_{12}$}

If we postulate that the Majorana mass matrix must have the $\mu-\tau$ symmetry, and apply the operation (\ref{diagmatrix}), then if we start with the matrix:
\begin{equation}
M_{M}^{(\mu \leftrightarrow \tau)}=\left(\begin{array}{ccc}
M_{ee} & M_{e\mu} & M_{e\mu} \\
M_{e\mu} & M_{\mu\mu} & M_{\mu\tau} \\
M_{e\mu} & M_{\mu\tau} & M_{\mu\mu} \end{array} \right).\label{majornanamass}
\end{equation}
After applying the operation the ``diagonal'' matrix will be:
\begin{equation}
M^{diag}= \label{diagmass1}
\end{equation}
\[\left(\begin{array}{ccc}\scriptscriptstyle{M_{ee}\cos^{2}\theta_{12} + (M_{\mu\mu}+M_{\mu\tau})\sin^{2}\theta_{12}+\sqrt{2}M_{e\mu}\sin2\theta_{12}} & \scriptscriptstyle{\sqrt{2}M_{e\mu}\cos2\theta_{12} + 1/2(M_{\mu\mu}+M_{\mu\tau}-M_{ee})\sin2\theta_{12}} & \scriptscriptstyle{0} \\
\scriptscriptstyle{\sqrt{2}M_{e\mu}\cos2\theta_{12} + 1/2(M_{\mu\mu}+M_{\mu\tau}-M_{ee})\sin2\theta_{12}} & \scriptscriptstyle{M_{ee}\sin^{2}\theta_{12} + (M_{\mu\mu}+M_{\mu\tau})\cos^{2}\theta_{12} - 2\sqrt{2}M_{e\mu}\cos\theta_{12}\sin\theta_{12}} & \scriptscriptstyle{0} \\
\scriptscriptstyle{0} & \scriptscriptstyle{0} & \scriptscriptstyle{M_{\mu\mu}- M_{\mu\tau}}\end{array} \right).
\]
In order to make this mass matrix really diagonal it is necessary to impose a value on the angle $\sin\theta_{12}$, an easy algebraic operation shows that:
\begin{equation}
\tan2\theta_{12}= \frac{\sqrt{8}M_{e\mu}}{-M_{\mu\mu}-M_{\mu\tau}+M_{ee}}.\label{weinang}
\end{equation}

For a Majorana mass matrix we will have in the most general case 8 parameters, this means that we need 8 different experiments that can provide us with values, that will able us to completely determine all the elements of the matrix, equation (\ref{weinang}) is the first equation that we need to do this. It is clear that since the tangent is a real number, the phases of the right handed part of this equation have to be eliminated. This implies that:
\begin{equation}
Im\left[M_{e\mu}\right]= Im \left[-M_{\mu\mu}-M_{\mu\tau}+M_{ee}\right].\label{imagtan}
\end{equation}
This equation gives a constriction that helps us to determine the parameters of the Majorana mass matrix.

\section{Value for the Majorana Masses}

Using (\ref{weinang}) in (\ref{diagmass1}) we can work out the values for the masses of the neutrino in the mass eigenstate, as a factor of the masses in the flavor state and the Weinberg angle and have equations that will work for any Majorana particle. These are:
\begin{equation}
M_{1}=M_{ee}+ \sqrt{2} M_{e \mu} \tan \theta_{12}\label{m1},
\end{equation}
\begin{equation}
M_{2}=M_{ee}- \sqrt{2} M_{e \mu} \cot \theta_{12}\label{m2},
\end{equation}
and
\begin{equation}
M_{3}= M_{\mu \mu}- M_{\mu \tau}.\label{m3}
\end{equation}
In general these masses are going to be complex. 

\vspace{0.3cm}

Making a little algebra we can get the values of the square difference of the masses, which as explain earlier are important quantities, since this are the values we can measure experimentally.
\begin{equation}
\Delta M_{\odot}^{2} \equiv \Delta M_{21}^{2}= \left|M_2\right|^2 - \left|M_1\right|^2,
\end{equation}
and
\begin{equation}
\Delta M_{\scriptstyle{ATM}}^{2} \equiv \left|\Delta M_{32}^{2} \right|= \left| \left|M_3\right|^2 - \left|M_2\right|^2\right|.
\end{equation}
In the case where there are no phases (real masses), this equations can be written as
\begin{equation}
\Delta M_{21}^{2}=(M_{ee}+ M_{\mu \mu}+ M_{\mu \tau})\sqrt{8(M_{e \mu})^{2}+ (-M_{ee}+ M_{\mu \mu}+ M_{\mu \tau})^{2}}\label{m21},
\end{equation}
\begin{equation}
\Delta M_{32}^{2}=-2(M_{e \mu})^{2}+\frac{1}{2}(M_{\mu \mu})^{2}-3M_{\mu \mu}M_{\mu \tau}+ \frac{1}{2}(M_{\mu \tau})^{2}- \frac{1}{2}(M_{e e})^{2}- \label{m32}.
\end{equation}

\[\frac{1}{2} (M_{\mu \mu}+ M_{\mu \tau}+M_{e e})\sqrt{8(M_{e \mu})^{2}+(M_{\mu \mu}+ M_{\mu \tau}-M_{e e})^{2}}.\]

Experimentally we can measure the square difference of the masses, and $M_{3}$ (The hierarchy), so we have four equations to determine the values of the masses in matrix (\ref{majornanamass}). In the limit were all elements of the matrix are real, then we will have 4 parameters, and 4 equation, so the system is completely determined. 

\chapter{A Model for Neutrino Mass in the Seesaw mechanism}

The goal for this chapter is to use the $\mu$-$\tau$ symmetry to reduce the number of parameters on the mass matrices even further, so we will be able determine as many parameters as possible. For this we will diagonalize the Majorana mass matrix, and leave a non-diagonal Dirac mass matrix with all the phases of the model included on it, then we will make all the possible phase transformations to reduce more phases, and finally we will use the see-saw mechanism to calculate the left neutrino mass matrix with as few parameters as possible.

\section{The Lagrangian for the Model}

Now we will build a model for the neutrino mass, for this we will start by using the following components (with their corresponding quantum numbers):

\vspace{.3cm}

\begin{tabular}{|c|ccc|r|}
	\hline
Component &  $SU(3)$    &   $SU(2)$  & $U(1)$   & Notation  \\
	\hline
Left handed doublet for leptons $L$   & 1 & 2 & -1 & (1,2,-1)\\
Right handed singlet for leptons $\ell_{R}$   & 1 & 1 & -2  & (1,1,-2)\\
Right handed neutrino $N_R$ & 1 & 1 & 0 & (1,1,0)\\
Higgs doublet $\phi$   & 1 & 2 & 1  & (1,2,1) \\
	\hline
\end{tabular}

Then we will use a base were the charged lepton mass and the interaction with the $W$ boson are diagonal, for this we can use $3 \times 3$ flavor transformations (in a similar way as chapter 4), explicitly:
\begin{equation}
\mathscr{L}= - \bar{\ell}_{L}M_{\ell}\ell_{R} -\frac{e}{\sqrt{2}\sin2\theta_{W}}\bar{\nu}_{L}\gamma^{\mu}\ell_{L}W_{\mu}^{+}+H.c 
\end{equation}
\[ =-\bar{\ell}_{L}U_{L}U_{L}^{\dagger}M_{\ell}U_{R}U_{R}^{\dagger}\ell_{R} -\frac{e}{\sqrt{2}\sin2\theta_{W}}\bar{\nu}_{L}U_{L}^{\dagger}\gamma^{\mu}U_{L}^{\dagger}\ell_{L}W_{\mu}^{+}+H.c \]
\[ =\bar{\ell}_{L}M_{\ell}^{Diag}\ell_{R} -\frac{e}{\sqrt{2}\sin2\theta_{W}}\bar{\nu}_{L}\gamma^{\mu}\ell_{L}W_{\mu}^{+}+H.c.\]
Where the transformations used are: 
\begin{equation}
\begin{array}{ccc}
\ell_{L} &  \rightarrow & U_{L}^{\dagger}\ell_{L} \\
\nu_{L} & \rightarrow & U_{L}^{\dagger}\nu_{L} \end{array}\} \hspace{.3cm} \textbf{L}\rightarrow U_{L}^{\dagger}\textbf{L}.
\end{equation}
\[\ell_{R} \rightarrow U_{R}^{\dagger}\ell_{R}.\]
such that $U_{L}^{\dagger}M_{\ell}U_{R}= M_{\ell}^{Diag}$.

Now we are left with two non diagonal terms in the Lagrangian (the Majorana and the Dirac Masses), only the Majorana mass term can be diaginalized, because trying to diagonalize the Dirac mass term will make a change of base in $\nu_{L}$ and will cause mixing in the interaction with the W boson. 
Let's remember that we have assumed $ \mu- \tau$ symmetry, so the two term Lagrangian for the neutrino mass will be:
\begin{equation}
\mathscr{L}_{mass}^{(\mu \leftrightarrow \tau)}= -\overline{\nu_{L}}M_{D}^{(\mu \leftrightarrow \tau)}\nu_{R} -\frac{1}{2}\overline{(\nu_{R})^{c}}M_{M}^{(\mu \leftrightarrow \tau)}\nu_{R} + H.c.
\end{equation}
where the masses $M_{D}^{(\mu \leftrightarrow \tau)}$ and $M_{M}^{(\mu \leftrightarrow \tau)}$ are the ones defined in (\ref{mmaj}) and (\ref{mdirac}).

Accordingly to chapter 4 we can use (\ref{UPMNS}) to diagonalize the Majorana matrix, so the Lagrangian will transform into:
\begin{equation}
\mathscr{L_{mass}}^{(\mu \leftrightarrow \tau)}= -\overline{\nu_{L}}M_{D}^{(\mu \leftrightarrow \tau)}U_{(\mu \leftrightarrow \tau)}U_{(\mu \leftrightarrow \tau)}^{\dagger}\nu_{R} -\frac{1}{2}\overline{(\nu_{R})^{c}}U_{(\mu \leftrightarrow \tau)}U_{(\mu \leftrightarrow \tau)}^{T}M_{M}^{(\mu \leftrightarrow \tau)}U_{(\mu \leftrightarrow \tau)}U_{(\mu \leftrightarrow \tau)}^{T}\nu_{R} + H.c
\end{equation}

\[= -\overline{\nu_{L}}M_{D}'\nu_{R} -\frac{1}{2}\overline{(\nu_{R})^{c}}M_{M}^{Diag}\nu_{R} + H.c.
\]
Where we have used the orthogonal transformation:
\begin{equation}
U_{(\mu \leftrightarrow \tau)}^{T}\nu_{R} \rightarrow \nu_{R}.
\end{equation}
And the transformations for the masses:
\begin{equation}
U_{(\mu \leftrightarrow \tau)}^{T}M_{M}^{(\mu \leftrightarrow \tau)}U_{(\mu \leftrightarrow \tau)}=M_{M}^{Diag}, \hspace{.4cm} M_{D}U_{(\mu \leftrightarrow \tau)}=M_{D}'\label{Trans}.
\end{equation}

It's important to note here that the Majorana mass $M_{M}^{Diag}$, is not only diagonal, but has real eigenvalues, to observe this explicitly let's remember that the mixing matrix can be multiplied by a diagonal matrix with three phases, so the diagonalization of the Majorana mass is:
\begin{equation}
U^{T}M_{M}U=(V K)^{T}M_{M}(V K)=K (V^{T}M_{M}V) K=K M_{M}^{Diag} K =
\end{equation}
\[\left(
\begin{array}{ccc}
e^{i\alpha} & 0 & 0 \\
0 & e^{i\beta}  & 0 \\
0 & 0 & e^{i\gamma} \end{array} 
\right)\cdot \left(
\begin{array}{ccc}
M_{1}e^{i\psi_{1}} & 0 & 0 \\
0 & M_{2}e^{i\psi_{2}}  & 0 \\
0 & 0 & M_{3}e^{i\psi_{3}} \end{array} 
\right)\cdot \left(
\begin{array}{ccc}
e^{i\alpha} & 0 & 0 \\
0 & e^{i\beta}  & 0 \\
0 & 0 & e^{i\gamma} \end{array} 
\right) =
\]
\[\left(
\begin{array}{ccc}
M_{1}e^{i(\psi_{1}+2\alpha)} & 0 & 0 \\
0 & M_{2}e^{i(\psi_{2}+2\beta)}  & 0 \\
0 & 0 & M_{3}e^{i(\psi_{3}+2\gamma)} \end{array} 
\right).
\]
The choice $\alpha=- \psi_{1}/2$, $\beta=- \psi_{2}/2$ and  $\gamma=- \psi_{3}/2$, will make real the eigenvalues of the Majorana mass, as previously noted, but this mean that the phases $\alpha$, $\beta$ and $\gamma$ will not be free in the future.
This will leave as the only non-diagonal term the Dirac mass $M_{D}'$.

\section{Seesaw with the Dirac mass}

Now, we will see the form of the Dirac mass matrix, starting by a general form of the 3 x 3 matrix with $\mu$-$\tau$ symmetry: 
\begin{equation}
M_{D}=\left(\begin{array}{ccc}
D_{ee} & D_{e\mu} & D_{e\mu} \\
D_{\mu e} & D_{\mu \mu} & D_{\mu \tau} \\
D_{\mu e} & D_{\mu \tau} & D_{\mu \mu} \end{array}\right).
\end{equation}

Let's remember that in general the elements of this matrix are complex, so in principle we have 5 real parameters and 5 phases in the matrix, leaving a total of ten independent parameters.

Next we will use the PMNS mixing matrix with $\mu$-$\tau$ symmetry, calculated in equation (\ref{UPMNS}) of chapter 5.
\begin{equation}
U=\left(\begin{array}{ccc}
e^{i\alpha_{1}}\cos \theta_{12} & -e^{i\alpha_{2}} \sin \theta_{12} & 0 \\
\frac{e^{i\alpha_{1}}\sin \theta_{12}}{\sqrt{2}} & \frac{e^{i\alpha_{2}} \cos \theta_{12}}{\sqrt{2}} & -\frac{e^{i \alpha_{3}}}{\sqrt{2}} \\
\frac{e^{i\alpha_{1}}\sin \theta_{12}}{\sqrt{2}} & \frac{e^{i\alpha_{2}} \cos \theta_{12}}{\sqrt{2}} & \frac{e^{i \alpha_{3}}}{\sqrt{2}} \end{array}\right).
\end{equation}
And make the explicit transformation for the Dirac mass term using $U_{PMNS}$ as $U_{(\mu \leftrightarrow \tau)}$ of equation (\ref{Trans}), we made this choice because in chapter 5 we have demonstrated that this matrix is the one that diagonalize $M_{M}$.
\begin{equation}
M_{D}'=M_{D}\cdot U= 
\end{equation}
\[\left(\begin{array}{ccc}
\scriptstyle{e^{i\alpha_{1}}(D_{ee}\cos \theta_{12} + \sqrt{2} MD_{e\mu} \sin \theta_{12})} & \scriptstyle{e^{i\alpha_{2}}(\sqrt{2}D_{e\mu}\cos \theta_{12} - D_{ee} \sin \theta_{12})} & \scriptstyle{0} \\
\scriptstyle{\frac{1}{2} e^{i\alpha_{1}}(2D_{\mu e}\cos \theta_{12} + \sqrt{2}(D_{\mu \mu}+ D_{\mu \tau}) \sin \theta_{12})} & \scriptstyle{\frac{1}{2} e^{i\alpha_{2}}(\sqrt{2}(D_{\mu \mu} + D_{\mu \tau})\cos \theta_{12} - 2D_{\mu e} \sin \theta_{12})} & \scriptstyle{\frac{e^{i \alpha_{3}}(-D_{\mu \mu} + D_{\mu \tau})}{\sqrt{2}}} \\
\scriptstyle{\frac{1}{2} e^{i\alpha_{1}}(2D_{\mu e}\cos \theta_{12} + \sqrt{2}(D_{\mu \mu}+ D_{\mu \tau}) \sin \theta_{12})} & \scriptstyle{\frac{1}{2} e^{i\alpha_{2}}(\sqrt{2}(D_{\mu \mu} + D_{\mu \tau})\cos \theta_{12} - 2D_{\mu e} \sin \theta_{12})} & \scriptstyle{\frac{e^{i \alpha_{3}}(D_{\mu \mu} - D_{\mu \tau})}{\sqrt{2}}} \end{array}\right).
\]
This matrix has the form:
\begin{equation}
M_{D}'=\left(\begin{array}{ccc}
D_{e1}e^{i\alpha} & D_{e2}e^{i\gamma} & 0 \\
D_{\mu 1}e^{i\beta} & D_{\mu 2}e^{i\omega} & -D_{\mu 3}e^{i\rho} \\
D_{\mu 1}e^{i\beta} & D_{\mu 2}e^{i\omega} & D_{\mu 3}e^{i\rho} \end{array}\right)\label{diracmass1},
\end{equation}
where the components $D_{mn}$ are real. This means that this matrix has 10 parameters.

\vspace{.3cm}

With out loss of generality we can make another transformation, but this time a phase transformation for each standard lepton, by using a 3 x 3 diagonal matrix with three different phases. We can make this transformation in the left and right handed charged lepton and in the left handed neutrino, in the following way:
\begin{equation}
\nu_{L}\rightarrow P\nu_{L}, \hspace{.3cm} \ell_{L,R}\rightarrow P\ell_{L,R}.
\end{equation}
Note that we can't make the transformation for the right handed neutrino wave function, this is because we have assumed that the Majorana mass term is diagonal with real positive eigenvalues, if we make this transformation, the matrix will still be diagonal, but will have complex values; mathematically this can be made, but for the model we are building we want all the phases to be in the Dirac mass matrix.

Finally the Dirac mass matrix will be:
\begin{equation}
M_{D}''=P\cdot M_{D}'=\left(\begin{array}{ccc}
D_{e1} e^{i\alpha + \delta_{1}} & D_{e2} e^{i\gamma + \delta_{1}} & 0 \\
D_{\mu 1} e^{i\beta + \delta_{2}} & D_{\mu 2} e^{i\omega + \delta_{2}} & -D_{\mu 3} e^{i\rho + \delta_{2}} \\
D_{\mu 1} e^{i\beta + \delta_{3}} & D_{\mu 2} e^{i\omega + \delta_{3}} & D_{\mu 3} e^{i\rho + \delta_{3}} \end{array} \right),
\end{equation}
where the matrix P is:
\begin{equation}
P=\left(\begin{array}{ccc}
e^{i \delta_{1}} & 0 & 0 \\
0 & e^{i \delta_{2}} & 0 \\
0 & 0 & e^{i \delta_{3}} \end{array}\right)
\end{equation}

These three phases have no physical meaning, they have being introduced in order to eliminate phases in one of the mass matrices, but once the choice have being made, the phases can´t be changed, so we will be able to eliminate some phases in the Dirac mass matrix, or in the seesaw neutrino mass (as it will be seeing further in this chapter)

To calculate the neutrino mass we will use the seesaw mechanism (\ref{seemass1})
\begin{equation}
m_{\nu}=M_{D}''\frac{1}{M_{M}}M_{D}''^{T}\label{seemass}
\end{equation}
and to make the calculations easier let's note that the matrix $P$, will make a transformation on the seesaw mass, in the following way:
\begin{equation}
m_{\nu}=M_{D}''\frac{1}{M_{M}}M_{D}''^{T}=(P M_{D}')\frac{1}{M_{M}}(PM_{D}')^{T}=P(M_{D}'\frac{1}{M_{M}}M_{D}'^{T})P=Pm_{\nu}'P\label{transP}
\end{equation}
This means we can make the calculations of the components of the seesaw mass and then just include the phases of the P matrix.

The neutrino mass can be calculated component by component with the formula:
\begin{equation}
(m_{\nu})_{\ell \ell'}=\sum_{a} (M_{D}')_{\ell a} \frac{1}{(M_{M})_{a}} (M_{D}')_{\ell' a}.
\end{equation}
The components of the neutrino mass are:
\begin{equation}
(m_{\nu})_{11}=\frac{e^{2i\alpha}(D_{e 1})^{2}}{M_{1}} + \frac{e^{2i\gamma}(D_{e 2})^{2}}{M_{2}}\label{mnu11},
\end{equation}
\begin{equation}
(m_{\nu})_{12}=(m_{\nu})_{21}=(m_{\nu})_{13}=(m_{\nu})_{31}=\frac{e^{i\alpha+i\beta}(D_{e 1}D_{\mu 1})}{M_{1}} + \frac{e^{i\gamma + i\omega}(D_{e 2}D_{\mu 2})}{M_{2}},
\end{equation}
\begin{equation}
(m_{\nu})_{22}=(m_{\nu})_{33}=\frac{e^{2i\beta}(D_{\mu 1})^{2}}{M_{1}} + \frac{e^{2i\omega}(D_{\mu 2})^{2}}{M_{2}}+ \frac{e^{2i\rho}(D_{\mu 3})^{2}}{M_{3}},
\end{equation}
\begin{equation}
(m_{\nu})_{23}=(m_{\nu})_{32}=\frac{e^{2i\beta}(D_{\mu 1})^{2}}{M_{1}} + \frac{e^{2i\omega}(D_{\mu 2})^{2}}{M_{2}}- \frac{e^{2i\rho}(D_{\mu 3})^{2}}{M_{3}}\label{mnu33}.
\end{equation}
Is important to remember that $M_{i}$ are the masses for the right handed neutrino in the mass eigenstate and the $(m_{\nu})_{ij}$ are the neutrino masses in the left handed flavor state.

\vspace{.3cm}
Now we want to sum this complex numbers so they will be only a real number with a complex phase, for this we have to remember that for a complex number:
\begin{equation}
a+ib=ce^{i\phi} \longrightarrow c=\sqrt{a^{2}+b^{2}}, \hspace{.4cm} \phi=\arctan\frac{b}{a} \label{complex}.
\end{equation}
In the case of two real numbers each with a phase, the sum can be made by transforming the exponential to a sine and cosine, so we have a number similar to (\ref{complex}) 
\[ Ae^{i\phi_{1}} + Be^{i\phi_{2}}=A(\cos \phi_{1}+i\sin \phi_{1})+B(\cos \phi_{2}+i\sin \phi_{2})= \]
\[(A\cos \phi_{1}+B\cos \phi_{2}) + i(A\sin \phi_{1}+B \sin \phi_{2})=Ce^{i\phi_{3}}, \]
So we will have:
\begin{equation}
C=\sqrt{(A\cos \phi_{1}+B\cos \phi_{2})^{2}+(A\sin \phi_{1}+B \sin \phi_{2})^{2}}= \label{complexsum1}
\end{equation}
\[\sqrt{A^{2}+B^{2}+2AB(\cos \phi_{1}\cos \phi_{2}+\sin \phi_{1}\sin \phi_{2})}\]
and
\begin{equation}
\phi_{3}=\arctan\frac{A\sin \phi_{1}+B \sin \phi_{2}}{A\cos \phi_{1}+B\cos \phi_{2}}.\label{complexsum2}
\end{equation}
Then the neutrino mass can be written as:
\begin{equation}
m_{\nu}= \left(\begin{array}{ccc}
m_{ee} e^{i\phi_{ee}}& m_{e\mu}e^{i\phi_{e\mu}} & m_{e\mu}e^{i\phi_{e\mu}} \\
m_{e\mu} e^{i\phi_{e\mu}}& m_{\mu \mu}e^{i\phi_{\mu \mu}} & m_{\mu \tau}e^{i\phi_{\mu \tau}} \\
m_{e\mu} e^{i\phi_{e\mu}}& m_{\mu \tau}e^{i\phi_{\mu \tau}} & m_{\mu \mu}e ^{i\phi_{\mu \mu}} \end{array}
\right). \label{M}
\end{equation}

With the phase transformation these components will be:
\begin{equation}
(m'_{\nu})_{11}=(m_{\nu})_{11}e^{2i\delta_{1}}\label{mnup11},
\end{equation}
\begin{equation}
(m'_{\nu})_{12}=(m'_{\nu})_{21}=(m_{\nu})_{12}e^{i\delta_{1}+ i\delta_{2}},
\end{equation}
\begin{equation}
(m'_{\nu})_{13}=(m'_{\nu})_{31}=(m_{\nu})_{13}e^{i\delta_{1}+ i\delta_{3}},
\end{equation}
\begin{equation}
(m'_{\nu})_{22}=(m_{\nu})_{22}e^{2i\delta_{2}},
\end{equation}
\begin{equation}
(m'_{\nu})_{23}=(m'_{\nu})_{32}=(m_{\nu})_{23}e^{i\delta_{2}+ i\delta_{3}},
\end{equation}
\begin{equation}
(m'_{\nu})_{33}=(m_{\nu})_{33}e^{2i\delta_{3}}\label{mnup33},
\end{equation}
To conserve symmetry we have to impose $\delta_{2}=\delta_{3}$. Which is logical and could have been done since the beginning in the matrix $P$, since different phases brake the symmetry as have been seen. The Neutrino mass matrix will be then:
\begin{equation}
m_{\nu}= \left(\begin{array}{ccc}
m_{ee} e^{i(\phi_{ee}+2\delta_{1})}& m_{e\mu}e^{i(\phi_{e\mu}+ \delta_{1}+ \delta_{3})} & m_{e\mu}e^{i(\phi_{e\mu}+ \delta_{1}+ \delta_{3})} \\
m_{e\mu} e^{i(\phi_{e\mu}+ \delta_{1}+ \delta_{3})}& m_{\mu \mu}e^{i(\phi_{\mu \mu}+2\delta_{3})} & m_{\mu \tau}e^{i(\phi_{\mu \tau}+2\delta_{3})} \\
m_{e\mu} e^{i(\phi_{e\mu}+ \delta_{1}+ \delta_{3})}& m_{\mu \tau}e^{i(\phi_{\mu \tau}+2\delta_{3})} & m_{\mu \mu}e ^{i(\phi_{\mu \mu}+2\delta_{3})} \end{array}
\right)\label{mnu}.
\end{equation}
Now we will make a choice in the two $\delta$ phases in order to reduce the number of free parameters in the model. The choice will be so the only phases remaining will be in the diagonal of the matrix, the appropriate choice is then:
\begin{equation}
\delta_{1}=-\phi_{e\mu} +\phi_{\mu \tau}/2, \hspace{.4cm} \delta_{3}=-\phi_{\mu \tau}/2.
\end{equation}

This will result in the following matrix:
\begin{equation}
m_{\nu}= \left(\begin{array}{ccc}
m_{ee} e^{i(\phi_{ee}-2\phi_{e\mu}+\phi_{\mu \tau})}& m_{e\mu} & m_{e\mu} \\
m_{e\mu} & m_{\mu \mu}e^{i(\phi_{\mu \mu}-\phi_{\mu \tau})} & m_{\mu \tau} \\
m_{e\mu} & m_{\mu \tau} & m_{\mu \mu}e ^{i(\phi_{\mu \mu}-\phi_{\mu \tau})} \end{array}\label{mnu2}
\right).
\end{equation}
This matrix has four independent real parameter (masses), and two independent complex phases, this is two less independent parameters than the Dirac mass matrix, that has five independent real parameter (masses), and three independent complex phases, this is because the two $\delta$ phases could had been chosen before to eliminate two of five complex phases.

Using the constriction (\ref{imagtan}) we can find an equation that relates the two phases:
\begin{equation}
m_{ee}\sin(\phi_{ee}-2\phi_{e\mu}+\phi_{\mu \tau})=-m_{\mu \mu}\sin(\phi_{\mu \mu}-\phi_{\mu \tau})\label{constriction}.
\end{equation}

Matrix (\ref{mnu2}) has 4 real parameters, so we could use the measured values for $\Delta m_{32}^{2}$, $\Delta m_{21}^{2}$, $m_{3}$ and $\tan \theta_{\odot}$ to completely determine the real parameters of the matrix, leaving only one free phase, because the two phases will be related by equation (\ref{constriction}) .

\vspace{.3cm}

Since $m_\nu$ is a  mass matrix for a Majorana particle we can use equations (\ref{m1})-(\ref{m32}), to calculate the values of the mass of the neutrinos in the mass eigenvalue state, by using the values calculated in equations (\ref{mnu11})-(\ref{mnu33}) and (\ref{mnup11})-(\ref{mnup33}). This will lead to:
\begin{equation}
m_{1}=\frac{e^{2i\alpha}(D_{e 1})^{2}}{2M_{1}}+\frac{e^{2i\beta}(D_{\mu 1})^{2}}{M_{1}}+\frac{e^{2i\gamma}(D_{e 2})^{2}}{2M_{2}}+\frac{e^{2i\omega}(D_{\mu 2})^{2}}{M_{2}}+
\end{equation}

\[\scriptstyle{\frac{\sqrt{8(e^{i(\alpha+\beta)}(D_{e 1}D_{\mu 1})M_{2}+e^{i(\gamma+\omega)}(D_{e 2}D_{\mu 2})M_{1})^{2}+(e^{2i\gamma}(D_{e 2})^{2}M_{1}-2e^{2i\omega}(D_{\mu 2})^{2}M_{1}+e^{2i\alpha}(D_{e 1})^{2}M{2}-2e^{2i\beta}(D_{\mu 1})^{2}M_{2})^{2}}}{2M_{1}M_{2}}},
\]
\begin{equation}
m_{2}=\frac{e^{2i\alpha}(D_{e 1})^{2}}{2M_{1}}+\frac{e^{2i\beta}(D_{\mu 1})^{2}}{M_{1}}+\frac{e^{2i\gamma}(D_{e 2})^{2}}{2M_{2}}+\frac{e^{2i\omega}(D_{\mu 2})^{2}}{M_{2}}-
\end{equation}
\[\scriptstyle{\frac{\sqrt{8(e^{i(\alpha+\beta)}(D_{e 1}D_{\mu 1})M_{2}+e^{i(\gamma+\omega)}(D_{e 2}D_{\mu 2})M_{1})^{2}+(e^{2i\gamma}(D_{e 2})^{2}M_{1}-2e^{2i\omega}(D_{\mu 2})^{2}M_{1}+e^{2i\alpha}(D_{e 1})^{2}M{2}-2e^{2i\beta}(D_{\mu 1})^{2}M_{2})^{2}}}{2M_{1}M_{2}}},
\]
\begin{equation}
m_{3}= \frac{2 D_{\mu 3}^{2}e^{2i\rho}}{M_{3}}\label{M3}.
\end{equation}

These values are very interesting; a quick interpretation will tell us that the hierarchy for the right handed neutrino will invert the hierarchy for the neutrino. For example if the right handed neutrino has a normal hierarchy (i.e $M_{3}>>M_{2}>>M_{1}$), then the neutrino will have an inverted hierarchy (i.e $m_{2}, m_{1}>>m_{3}$), if in opposite the right handed neutrino have an inverted hierarchy  (i.e $M_{3}<<M_{2}<<M_{1}$), then the neutrino will likely have an normal hierarchy (i.e $m_{1}<<m_{2}<<m_{3}$). This will be further exemplify in chapter 8.

Another interesting fact that can be concluded of this equations is that mass of the first two neutrinos depend on a combination of the masses of the first two right handed neutrinos and four elements of the Dirac mass, but the mass of the third neutrino depends only on the mass of the third right handed neutrino, and one element of the Dirac mass.
\chapter{Leptogenesis}

In this chapter we will introduce the concept of baryon asymmetry in the universe and use the explanation of baryogenesis via leptogenesis to calculate the baryon asymmetry parameter. We will have one more equations to use for determine a parameter in the model, so far.

\section{Leptogenesis and Sakharov Conditions}
If we assume that immediately after the Big Bang there was the same quantity of matter and antimatter in the universe (a fair assumption we believe), then we encounter with a big problem, one that has been around for more than half a century, and has been referred in cosmology as the matter-antimatter asymmetry, which as the name explains, simply refers to the observation that the universe shows much more matter present than antimatter. Even worse, almost all antimatter observed can be attribute to collisions between primary particles and interstellar medium, this means it was not created in the Big Bang.

\vspace{.3cm}

Neutrinos were extremely important in the early universe; we believe that a large quantity was created in those first moments, and that still today neutrinos form a homogeneous background in all the universe (approximately 300 neutrinos per $cm^{3}$), similar to the microwave background. This means that the mass of the neutrino is an important parameter for the evolution of the universe and played a role in the formation of the supercumulus of galaxies. Basically the physical reactions in the first minutes of the universe are the same as in the center of the stars, the primordial reaction is two protons (hydrogen nucleus) fusing to become a helium atom, and then this reactions continue to generate heavier atoms. When this happened in the early universe is what we referred as nucleosynthesis: The composition of the universe was ``decided'' in this moments which lasted only three minutes. The exact prediction of the abundance of elements at this time depends on the baryon-antibaryon asymmetry and on the baryonic excess of photons (radiation).    

Using the WMAP observation \cite{Spergel}, we now know the baryon-antibaryon asymmetry of the universe is given by:
\begin{equation}
Y_{B}\equiv \frac{\textsl{n}_{B}-\textsl{n}_{\bar{B}}}{s} \approx \frac{\eta_{B}}{7.04} \approx (8.66\pm.28)\times 10^{-11}.
\end{equation}
Here $s$ is the entropy density and $\eta_{B}$ is the baryon to photon radio
\begin{equation}
\eta_{B}=n_{B}/n_{\gamma}=(6.0\pm0.2)\times10^{-10}.
\end{equation}
This number is extremely interesting, it actually means that right after Big Bang nucleosynthesis there was ten billion and one quarks for every ten billion antiquarks, so after annihilation one quark survived, and all the antimatter was eliminated.  

\vspace{.3cm}

A solution to this problem was proposed by Sakharov in 1967 \cite{Sakharov}, he stated that there are three conditions that have to be met simultaneously in order to cause baryon asymmetry if we start with a baryon symmetric universe.
\begin{enumerate}
	\item Baryon number violation ($\Delta B\neq 0$)
	\item C and CP violation
	\item A departure from thermal equilibrium (baryon number violating processes out of equilibrium)
\end{enumerate}

The second condition has been seen in particle decays, such as $K_{L}^{0}\rightarrow 2\pi$  and  $B_{d}\rightarrow J/\psi K_{s}^{0}$ for example.
The first condition is allowed in the Standard Model, 't~Hooft sphaleron's \cite{Hooft}, that are a solutions in the SM at non perturbative level, that show that there is a $B - L$ symmetry, this means that even if $B$ and $L$ are not conserve separately (at one loop level the baryon and lepton number currents have anomalies) $B - L$ is. This is due to an infinite degeneration of the vacuum (minimum energy state), in non-abelian gauge theory. The transition between this states is by means of tunneling with an exponentially decreasing factor that suppress this process, though making the baryon asymmetry very small. In order for a sphaleron process to be viable as a mechanism for baryon asymmetry it is necessary to have very high temperatures (close to the electroweak phase transition) \cite{KRS}, but this will require that the Higgs boson mass have an upper bound of $m_{H}<80$ GeV \cite{cohen}, since the lower limit measured by experiments for this mass is $m_{H}>114$ GeV this process can't explain the asymmetry parameter, so it will be necessary to have another logical explanation. 

\vspace{.3cm}

In 1986 Fukugita and Yanagida proposed a baryogenesis via leptogenesis model \cite{Fukugita}, in this model the right handed neutrino of the seesaw mechanism decayed in the early universe producing the leptons of the SM, and the Higgs boson; if there is CP violation in this process, then, there will be a lepton-antilepton number violation, which lead to a lepton excess in the universe, and then via a spheralon interaction this lepton number violation will convert into a baryon number since we require a $B - L$ conservation. So, if in the seesaw neutrino mass matrix, there are CP violating phases, we will have leptogenesis, and might be able to explain the origin of matter abundance (it is important to note that in order to have creation of lepton via the decay of a heavy neutral lepton it is necessary that the decay rate be slower than the Hubble expansion rate, because this will create a suppressed back reaction).

\section{Baryon Asymmetry Parameter}

To have leptogensis we need to have thermal disequilibrium, so this process have to take place in a part of the universe were the neutrino reactions are below the Hubble expansion rate, and then the process will start at a temperature $T$, were $T\sim M_{1}$, when $M_{1}$ is the lightest of the heavy neutral leptons. This means that the other two heavy neutral leptons have disappeared via decays, because at higher temperatures they were in equilibrium, but for a temperature $T$ they are not. Since we only have now the lightest of the heavy neutral leptons ($N_{1}$), then only this particle is the responsible for leptogenesis and consequently lepton asymmetry. The lepton asymmetry parameter will be then: 
\begin{equation}
\epsilon_{1}=\frac{\Gamma(N_{1}\rightarrow\ell\phi)-\Gamma(N_{1}\rightarrow\bar{\ell}\phi)}{\Gamma(N_{1}\rightarrow\ell\phi)+\Gamma(N_{1}\rightarrow\bar{\ell}\phi)}.
\end{equation}
Where $\ell$ is the lepton doublet and $\phi$ is the Higgs doublet.

\vspace{.3cm}

The baryon asymmetry is proportional to the parameter $\epsilon_1$, but it is necessary to include a constant in order to have an equality, 
\begin{equation}
\frac{n_B}{n_\gamma} \approx \frac{ \epsilon_{1} \eta}{g}.
\end{equation} 
The constant $\eta$ is an efficiency factor, it has to be calculated taking into account how much out of equilibrium the heavy neutral lepton decay is (remember that out of equilibrium decay is the third of the  Sakharov conditions). The other constant introduced is due to the fact that only $N_1$ is responsible for the baryon asymmetry, so we divide by the spin degrees of freedom of the particles in the Standard Model ($g=118$). The baryon asymmetry parameter $\epsilon_{1}$ is in the order of $10^{-6}$.

\begin{figure}[ht]
\centerline{
\epsfxsize=400pt
\epsfbox{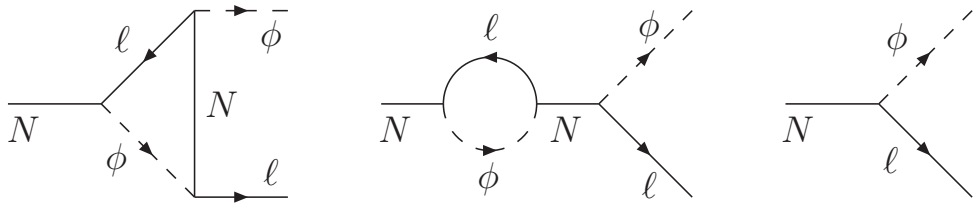}}
\caption{\em Diagrams contributing to CP-violating $N$ decay.\label{fig:Ndecay}}
\end{figure}

The tree-level decay width of $N_1$ is  $ \Gamma(N_{1}\rightarrow\ell\phi)=\Gamma(N_{1}\rightarrow\bar{\ell}\phi) = {\left|h _{1b}\right|^2 M_1}/{8\pi} $.
So at tree level the decay rate to leptons and to antileptons is the same. Using the interference between tree and loop diagrams shown in fig. 7.1 we can calculate the decay rates:
\begin{equation}
\Gamma(N_1\to \ell \phi) \propto \sum_{c} \left|h_{1c} + A h_{1b} h_{db}^{*} h_{dc} \right|^{2}, \qquad
\Gamma(N_1\to \bar{\ell} \phi) \propto \sum_{c} \left|h^{*}_{1c} + A h_{1b}^{*} h_{db} h_{dc}^{*}\right|^2.
\end{equation}
where $A$ is the complex loop factor. With these decay rates we can calculate the lepton asymmetry parameter:
\begin{equation}
\epsilon_{1}=\frac{1}{8\pi(hh^{\dagger})_{11}}\sum_{b=2}^{3} Im\left[(hh^{\dagger})_{1b}^{2}\right] f(\frac{M_{b}^{2}}{M_{1}^{2}})
\end{equation}
Where the function is:
\begin{equation}
f(x)=\sqrt{x}\left[(1+x)\ln(\frac{x}{1+x})+\frac{2-x}{1-x},\right]
\end{equation}
here, $h_{ij}$ is the Yukawas related to the Dirac mass, and $M_{i}$ the masses of the right handed neutrinos.

In the specific case were $M_{3}>>M_{1}$ and $M_{2}>>M_{1}$ then $f(x)\approx -3/2\sqrt{x}$ the parameter $\epsilon_{1}$ will be calculated as:
\begin{equation}
\epsilon_{1}=-\frac{3}{16\pi(hh^{\dagger})_{11}}\sum_{b=2}^{3} Im\left[(hh^{\dagger})_{1b}^{2}\right] \frac{M_{1}}{M_{b}} \label{eps1}.
\end{equation}
We can actually assume that the mass hierarchy is different, since we don't actually know the mass of the heavy leptons, by this we mean that the mass $M_{2}$ or $M_{3}$, could be the lightest. So we can define the parameters $\epsilon_{2}$ and $\epsilon_{3}$  which will required only to change some of the indices.

\vspace{.3cm}

To calculate explicitly $\epsilon_{1}$, first let's remember that $v h=M$, where $v$ is the vacuum expectation value and M is the Dirac mass matrix, next we calculate the multiplication of the Yukawas with it's conjugate. For the denominator we will use

\begin{equation}
(hh^{\dagger})_{11}=\left|h_{11}\right|^{2}+\left|h_{12}\right|^{2}+\left|h_{13}\right|^{2} ,
\end{equation}
for the numerator we will have to calculate the imaginary part of
\begin{equation}
(hh^{\dagger})_{12}=\left|h_{11}\right|\left|h_{21}\right|e^{i(\alpha_{11}-\alpha_{21})}+\left|h_{12}\right|\left|h_{22}\right|e^{i(\alpha_{12}-\alpha_{22})}+\left|h_{13}\right|\left|h_{23}\right|e^{i(\alpha_{13}-\alpha_{23})},
\end{equation}
and
\begin{equation}
(hh^{\dagger})_{13}=\left|h_{11}\right|\left|h_{31}\right|e^{i(\alpha_{11}-\alpha_{31})}+\left|h_{12}\right|\left|h_{32}\right|e^{i(\alpha_{12}-\alpha_{32})}+\left|h_{13}\right|\left|h_{33}\right|e^{i(\alpha_{13}-\alpha_{33})}.
\end{equation}
Replacing the values of the matrix (\ref{diracmass1}), the parameter $\epsilon_{1}$ is:

\begin{equation}
\scriptstyle{\epsilon_{1}=-\frac{3}{16\pi v^{2}(D_{e 1}^{2}+D_{e 2}^{2})}(D_{e 1}^{2}D_{\mu 1}^{2}\sin (2(\alpha-\beta))+D_{e 2}^{2}D_{\mu 2}^{2}\sin (2(\gamma-\omega))+2D_{e 1}D_{\mu 1}D_{e 2}D_{\mu 2}\sin (\alpha+\gamma-\beta-\omega))(\frac{M_{1}}{M_{2}}+\frac{M_{1}}{M_{3}})}\label{epsp}.
\end{equation}

This formula depends on four real parameters and two relative phases. We have to remember that at this point the real parameters for the Majorana mass matrix can be determinated by the values of the solar angle $(\theta_{\odot})$ and the square difference of the neutrino mases ($\Delta m_{\odot}^2$ and $\Delta m_{ATM}^2$) up to a phase contribution. This will leave 4 equations and 5 variables for the Dirac mass matrix real parameters, and three complex phases;  by invoking the measured value of $\epsilon_1$, we can determine one more phase. This will leave only one free phase at low energies in the complete model, and three free parameters at high energies (one mass and two phases).

\chapter{An Example of the Model}

In this part of the thesis we will like to exemplify a particular case. In this particular example we will take the Majorana masses and give them values, this values have actually not been measured, but we feel that making this assumptions will let the model be further understood.
The values we will take are $M_{3}= 10^{15} GeV$, $M_{2}= 10^{12} GeV$ and  $M_{1}= 10^{9} GeV$. Or simply put, we will make an approximation taking $M_{3}>>M_{2}>>M_{1}$.

\vspace{.3cm}

Using this values an making approximations, the first thing we can see is that the values for the neutrino masses in the seesaw model are:
\begin{equation}
(M_{\nu})_{11}=\frac{e^{2i\alpha}(D_{e 1})^{2}}{M_{1}}\label{m11},
\end{equation}
\begin{equation}
(M_{\nu})_{12}=(M_{\nu})_{21}=(M_{\nu})_{13}=(M_{\nu})_{31}=\frac{e^{i\alpha+i\beta}(D_{e 1}D_{\mu 1})}{M_{1}} \label{m12},
\end{equation}
\begin{equation}
(M_{\nu})_{22}=(M_{\nu})_{33}=(M_{\nu})_{23}=(M_{\nu})_{32}=\frac{e^{2i\beta}(D_{\mu 1})^{2}}{M_{1}} \label{m22}.
\end{equation}

And making the phase transformation (\ref{transP}), then we will have the following elements:
\begin{equation}
(M_{\nu})_{11}=\frac{e^{2i\alpha+2i\delta_{1}}(D_{e 1})^{2}}{M_{1}} ,
\end{equation}
\begin{equation}
(M_{\nu})_{12}=(M_{\nu})_{21}=(M_{\nu})_{13}=(M_{\nu})_{31}=\frac{e^{i\alpha+i\beta+i\delta_{1}+ i\delta_{3}}(D_{e 1}D_{\mu 1})}{M_{1}} ,
\end{equation}
\begin{equation}
(M_{\nu})_{22}=(M_{\nu})_{33}=(M_{\nu})_{23}=(M_{\nu})_{32}=\frac{e^{2i\beta+2i\delta_{3}}(D_{\mu 1})^{2}}{M_{1}} ,
\end{equation}
Which means that the simple choice $\alpha=-\delta_{1}$ and $\beta=-\delta_{3}$ will completely remove all the complex phases in the neutrino mass matrix, leaving it in the form:
\begin{equation}
M_{\nu}=\left(\begin{array}{ccc}
m_{ee} & m_{e\mu} & m_{e\mu} \\
m_{e\mu} & m_{\mu\mu} & m_{\mu\tau} \\
m_{e\mu} & m_{\mu\tau} & m_{\mu\mu} \end{array} \right)=\left(\begin{array}{ccc}
\frac{(D_{e 1})^{2}}{M_{1}}  & \frac{(D_{e 1}D_{\mu 1})}{M_{1}}  & \frac{(D_{e 1}D_{\mu 1})}{M_{1}} \\
\frac{(D_{e 1}D_{\mu 1})}{M_{1}} & \frac{(D_{\mu 1})^{2}}{M_{1}} & \frac{(D_{\mu 1})^{2}}{M_{1}} \\
\frac{(D_{e 1}D_{\mu 1})}{M_{1}} & \frac{(D_{\mu 1})^{2}}{M_{1}} & \frac{(D_{\mu 1})^{2}}{M_{1}}\end{array} \right).
\end{equation}
Then the Dirac mass matrix with this choice of phases becomes:
\begin{equation}
M_{D}=\left(\begin{array}{ccc}
D_{e1} & D_{e2} e^{i(\gamma')} & 0 \\
D_{\mu 1} & D_{\mu 2} e^{i(\omega')} & -D_{\mu 3} e^{i(\rho')} \\
D_{\mu 1} & D_{\mu 2} e^{i(\omega')} & D_{\mu 3} e^{i(\rho')} \end{array} \right),
\end{equation}
where $\gamma'=\gamma-\alpha$ $\omega'=\omega-\beta$ and $\rho'=\rho- \beta$. Also it can be seen that this matrix have 5 real masses and three phases, as it was seen in chapter 6.

\vspace{.3cm}

With this information now we can make predictions for the values of some of the parameters of the model, for this we have to remember that in (\ref{massmuitau}) we have equations that give the elements of the neutrino mass in the flavor eigenstate as a factor of the masses in the mass eigenstate and the solar angle. Since $m_3$ is zero $m_{\mu \mu} =m_{\mu \tau}$, all the elements can be calculated with the equations:
\begin{equation}
m_{ee}=m_1\cos^{2}\theta_{\odot} + m_2\sin^{2}\theta_{\odot},
\end{equation}
\begin{equation}
m_{e\mu}=(m_1-m_2)\cos\theta_{\odot}\sin\theta_{\odot}/\sqrt{2},
\end{equation}
\begin{equation}
m_{\mu\mu}=m_{\mu \tau}=(m_2\cos^{2}\theta_{\odot}+m_1\sin^{2}\theta_{\odot})/2.
\end{equation}

We can use the experimentally measured values of the neutrino solar and atmospheric oscillations, to calculate the values of the masses $m_1$ and $m_2$, remembering that the mass $m_3$ is zero.
\begin{equation}
\Delta m_{ATM}^2=m_2^2 \rightarrow m_2=\pm\sqrt{\Delta m_{\scriptstyle{ATM}}^2},
\end{equation}
\begin{equation}
\Delta m_{\odot}^2=m_2^2-m_1^2 \rightarrow m_1=\pm\sqrt{\Delta m_{\scriptstyle{ATM}}^2-\Delta m_{\odot}^2}.
\end{equation}
We will have two cases, for the sign of the masses (the other two possible cases are just a $\pi/2$ phase transformation on the mass elements in the Lagrangian):

\vspace{.3cm}

i) $m_1 > 0$, $m_2 >0$

In this case, the three elements that we can calculate for the mass matrix, in the flavor base are:
\begin{equation}
m_{ee}=\sqrt{\Delta m_{\scriptstyle{ATM}}^2-\Delta m_{\odot}^2}\cos^{2}\theta_{\odot} + \sqrt{\Delta m_{\scriptstyle{ATM}}^2}\sin^{2}\theta_{\odot}\label{mee1},
\end{equation}
\begin{equation}
m_{e\mu}=(\sqrt{\Delta m_{\scriptstyle{ATM}}^2-\Delta m_{\odot}^2}-\sqrt{\Delta m_{\scriptstyle{ATM}}^2})\cos\theta_{\odot}\sin\theta_{\odot}/\sqrt{2},
\end{equation}
\begin{equation}
m_{\mu \mu}=\sqrt{\Delta m_{\scriptstyle{ATM}}^2}\cos^{2}\theta_{\odot} + \sqrt{\Delta m_{\scriptstyle{ATM}}^2-\Delta m_{\odot}^2}\sin^{2}\theta_{\odot}.
\end{equation}

ii) $m_1 > 0$, $m_2 <0$

In this other case the corresponding elements are:
\begin{equation}
m_{ee}=\sqrt{\Delta m_{\scriptstyle{ATM}}^2-\Delta m_{\odot}^2}\cos^{2}\theta_{\odot} - \sqrt{\Delta m_{\scriptstyle{ATM}}^2}\sin^{2}\theta_{\odot}\label{mee2},
\end{equation}
\begin{equation}
m_{e\mu}=(\sqrt{\Delta m_{\scriptstyle{ATM}}^2-\Delta m_{\odot}^2}+\sqrt{\Delta m_{\scriptstyle{ATM}}^2})\cos\theta_{\odot}\sin\theta_{\odot}/\sqrt{2},
\end{equation}
\begin{equation}
m_{\mu \mu}=\sqrt{\Delta m_{\scriptstyle{ATM}}^2}\cos^{2}\theta_{\odot} - \sqrt{\Delta m_{\scriptstyle{ATM}}^2-\Delta m_{\odot}^2}\sin^{2}\theta_{\odot}.
\end{equation}

The experientially measured values for the solar angle and the atmospheric and solar oscillation masses are: $\theta_{\odot}=33.7 \pm 1.3$, $\Delta m_{\scriptstyle{ATM}}^2= 2.6 \pm .2 \times 10^{-3} eV^2$ , $\Delta m_{\odot}^2= 7.9 \pm .28 \times 10^{-5} eV^2$. With this values we calculate for the two cases: 

i) $m_{ee}=0.05 \pm 0.002$ eV, $m_{e \mu}=-0.00025 \pm 0.000014$ eV, $m_{\mu \mu}=0.051 \pm 0.002$ eV.

\vspace{.3cm}

ii) $m_{ee}=0.019 \pm 0.00225$ eV, $m_{e \mu}=0.033 \pm 0.0014$ eV, $m_{\mu \mu}=-0.0198 \pm 0.002$ eV.

The calculated value for $m_{ee}$ is a very important part of the model, since it can be measured experimentally via the neutrinoless double beta decay. The accepted upper limit for this mass is $<(0.41-0.98)$ \cite{Avignone}  which is well above the value calculated via the equations (\ref{mee1}) and (\ref{mee2}).

\vspace{.3cm}

The two elements in the first column of the Dirac mass matrix can be calculated using:
\begin{equation}
D_{e1}=\pm\sqrt{m_{ee}M_1}\label{de1},
\end{equation}
and
\begin{equation}
D_{\mu 1}= \pm\sqrt{m_{\mu \mu}M_1}\label{dmu1}
\end{equation}

We can also use equation (\ref{eps1}) to find a similar result as (\ref{epsp}):

\begin{equation}
\scriptstyle{\epsilon_{1}=-\frac{3}{16\pi v^{2}(D_{e 1}^{2}+D_{e 2}^{2})}(D_{e 2}^{2}D_{\mu 2}^{2}\sin (2(\gamma'- \omega'))+2D_{e 1}D_{\mu 1}D_{e 2}D_{\mu 2}\sin (\gamma'-\omega'))(\frac{M_{1}}{M_{2}})}.
\end{equation}
This equation depends in only one relative phase and four parameters (where two have already been determined), so we can use it to eliminate one more free parameter.

\vspace{.3cm}

At the end by making approximations using $M_{3}>>M_{2}>>M_{1}$, we have completely determine the neutrino mass at low energies (using see-saw), but have left two undetermined real masses at high energies (Dirac mass matrix). We have also make an important prediction  ($m_{ee}$) that can eventually (with more sensitive experiments) tell if the model, or the symmetry used, are close to reality.

\vspace{.3cm}

Some important facts about the approximation: First, the mass $m_3$ is not actually zero, if we look at equation (\ref{M3}), we realize that a high value of $M_3$ is in the denominator will lead to a low value for $m_3$. If all the Dirac masses are in the same order of magnitude, then the order difference between $M_1$ and $M_3$ will be the same as $m_3$ and $m_1$. Secondly, the values  calculated for $m_{ee}$, $m_{e \mu}$ and $m_{\mu \mu}$, only depend on the fact that $m_{\mu \mu}\approx m_{\mu \tau}$, this will leave only three parameters in the matrix, being completely determinated. Note that for this to be possible is not necessary that $m_3 =0$, only that $m_3 << \sqrt{\Delta m_{\odot}^2}$. Finally, a quick observation of equations (\ref{m11})-(\ref{m22}), will lead to the identity $m_{ee}m_{\mu \mu}\approx m_ {e \mu}^2$, this relation does not show in the calculated values for these masses, this means that the hierarchy has to be soft, i.e. the values between $M_1$ and $M_2$ can't be that far apart. This will leave to more terms added in equations (\ref{m11})-(\ref{m22}) that inversely depend on $M_2$, which eventually means that we will not be able to eliminate all the phases at low energies, although their contribution to above estimations is negligible. 

\chapter{Conclusions and Perspectives for the Future}

The motivation for this thesis was to construct a neutrino model were all the elements of the mass matrix are determinated, the reality is that after making all the calculation we realized that this objective was not met. Experimentally there are six quantities we are able to measure: two square mass differences, three angles, and the mass hierarchy. The values for two angles where used in the mixing matrix, leaving four quantities that we can use to try to completely determine the parameters of the mass matrix. 

At low energies the neutrino mass matrix has six parameters: four real masses, and two complex phases. So we can determine the four masses, but we will still have one CP violation phase that will be a free parameter for the model (we must remember that there are two CP phases, but they are related by a constriction due to the fact that we require that the solar angle is real). On the other hand, for high energies, we have five masses and three phases, in principle we will have one undeterminated mass and three phases, but one of the phases can be fix via the baryon asymmetry parameter, so there are two undeterminated parameters (one mass and two phases related by a constriction). So there is loss of information (two parameters, that become only one after using the baryon asymmetry parameter) when we pass form high energies to low energies. Also, there are three mases for the right handed neutrinos that are unknow, using different values for them will lead to predictions at low energies, comparing the predictions with experimental values may give some information about the actual value of the right handed neutrinos. An example of this was given in chapter eight, we assumed that $M_3$ is much higher than the other two, this lead to a low value for $m_3$ and to numeric values for the Majorana mass terms, comparing this to the upper limit experimentally measured for the neutrinoless double beta decay ($m_{ee}$) suggests that the value chosen for $M_3$ might be in the right order of magnitude.

We are very interested in having a completely determinated model with out any free parameters, this can be accomplished in two different ways, giving more physical arguments, that will reduce the parameters in the Dirac mass matrix, or inventing more experiments that will give more information, for example there are experiments trying to measure CP violation at low energies. These experiments (if successful), will measure one CP violation phase, and then we could completely determine the model at low energies.

Other interesting result for the model is that, given that we have CP violation phases in the Dirac mass, we have enough CP violation to consider baryogenesis via leptogenesis to be a viable explication for the baryon asymmetry in th universe.

One last important thing to note, is that the $\mu-\tau$ is not an exact symmetry. In the charged leptons, the symmetry does not exist, the mass for the neutrino has to be different, which can be calculated by using perturbations at one loop. However the violation of $\mu-\tau$ symmetry induced by such a source is very small \cite{juan}.

\vspace{.4cm}

Finally, some perspectives of what more can be done following the premises of the thesis.
\begin{enumerate}
	\item{Instead of using the hierarchy for the right handed masses where $M_3$ is the heaviest, we can use a different hierarchy, that will lead to not only different surviving terms in an approximation similar to the one in chapter eight, but will also lead to different baryon asymmetry parameters, depending on the lightest of the masses.}
	\item{There are some interesting theories that propose neutrinos as dark matter. One of them is adding sterile neutrinos to the model. Other is, that if the right handed neutrinos have much lower mass than the ones used in this thesis ($\sim 100$ KeV), they can generate dark matter (but there can not be baryogenesis via leptogenesis). Both of this theories can be explored using $\mu-\tau$ symmetry,}
	\item{$\mu-\tau$ symmetry can be introduce in other models, for example $SU(2)_{R} \times SU(2)_{L} \times U(1)_{B-L}$. Where now the right handed neutrinos are part of a doublet joint to the handed  charged leptons. The symmetry can be used in other high dimension and unification models.}
	\item{The left Majorana mass term that was not used for simplicity, can be introduced in the model in order to do seesaw type II.}
	\item{Neutrino measurements and theories are actually very useful in cosmology to explain, and measure, processes as stellar reactions, supernovas, and even the Big Bang. Observation of the universe can help to measure more parameters that can be used in the theory.}
\end{enumerate}

\backmatter
\chapter*{Appendix A. Explicit values for Majorana masses}

Making the sum of complex numbers (\ref{complexsum1}) and (\ref{complexsum2}) on equations (\ref{mnu11})-(\ref{mnu33}), will lead to an the following components for matrix (\ref{M}):

\begin{equation}
(m_{\nu})_{11}=(m_{\nu})_{ee}e^{\phi_{ee}}=
\end{equation}
\[\scriptstyle{\sqrt{\frac{(D_{e 1})^{4}}{(M_{1})^{2}}+\frac{(D_{e 2})^{4}}{(M_{2})^{2}}+\frac{2(D_{e 1})^{2}(D_{e 2})^{2} \cos\left[2(\alpha-\gamma)\right]}{M_{1}M_{2}}}} \ast e^{2i\delta_{1}+i\arctan\left[\frac{M_{2}(D_{e 1})^{2}\sin(2\alpha)+M_{1}(D_{e 2})^{2}\sin(2\gamma)}{M_{2}(D_{e 1})^{2}\cos(2\alpha)+M_{1}(D_{e 2})^{2}\cos(2\gamma)}\right]}\]

\begin{equation}
(m_{\nu})_{12}=(m_{\nu})_{21}=(m_{\nu})_{13}=(m_{\nu})_{31}=(m_{\nu})_{e\mu}e^{\phi_{e\mu}}=
\end{equation}
\[\scriptstyle{\sqrt{\frac{(D_{e 1})^{2}(D_{\mu 1})^{2}}{(M_{1})^{2}}+\frac{(D_{e 2})^{2}(D_{\mu 2})^{2}}{(M_{2})^{2}}+\frac{2D_{e 1}D_{e 2}D_{\mu 1}D_{\mu 2} \cos\left[\alpha+\beta-\gamma-\omega\right]}{M_{1}M_{2}}}} \ast\]
\[ e^{i(\delta_{1}+\delta_{3})+i\arctan\left[\frac{M_{2}D_{e 1}D_{\mu 1}\sin(\alpha+\beta) + M_{1}D_{e 2}D_{\mu 2}\sin(\gamma+\omega)}{M_{2}D_{e 1}D_{\mu 1}\cos(\alpha+\beta) + M_{1}D_{e 2}D_{\mu 2}\cos(\gamma+\omega)}\right]}\]

\begin{equation}
(m_{\nu})_{22}=(m_{\nu})_{33}=(m_{\nu})_{\mu \mu}e^{\phi_{\mu \mu}}=
\end{equation}
\[\scriptstyle{\sqrt{\frac{(D_{\mu 1})^{4}}{(M_{1})^{2}}+\frac{(D_{\mu 2})^{4}}{(M_{2})^{2}}+\frac{(D_{\mu 3})^{4}}{(M_{3})^{2}}+\frac{2(D_{\mu 1})^{2}(D_{\mu 3})^{2} \cos\left[2(\beta-\rho)\right]}{M_{1}M_{3}}+\frac{2(D_{\mu 1})^{2}(D_{\mu 2})^{2} \cos\left[2(\beta-\omega)\right]}{M_{1}M_{2}}+\frac{2(D_{\mu 2})^{2}(D_{\mu 3})^{2} \cos\left[2(\rho-\omega)\right]}{M_{2}M_{3}}}} \ast \]
\[ e^{2i\delta_{3}+i\arctan\left[\frac{M_{2}M_{3}(D_{\mu 1})^{2}\sin(2\beta)+M_{1}M_{2}(D_{\mu 3})^{2}\sin(2\rho)+M_{1}M_{3}(D_{\mu 2})^{2}\sin(2\omega)}{M_{2}M_{3}(D_{\mu 1})^{2}\cos(2\beta)+M_{1}M_{2}(D_{\mu 3})^{2}\cos(2\rho)+M_{1}M_{3}(D_{\mu 2})^{2}\cos(2\omega)}\right]}
\]

\begin{equation}
(m_{\nu})_{23}=(m_{\nu})_{32}=(m_{\nu})_{\mu \tau}e^{\phi_{\mu \tau}}=
\end{equation}
\[\scriptstyle{\sqrt{\frac{(D_{\mu 1})^{4}}{(M_{1})^{2}}+\frac{(D_{\mu 2})^{4}}{(M_{2})^{2}}+\frac{(D_{\mu 3})^{4}}{(M_{3})^{2}}-\frac{2(D_{\mu 1})^{2}(D_{\mu 3})^{2} \cos\left[2(\beta-\rho)\right]}{M_{1}M_{3}}+\frac{2(D_{\mu 1})^{2}(D_{\mu 2})^{2} \cos\left[2(\beta-\omega)\right]}{M_{1}M_{2}}-\frac{2(D_{\mu 2})^{2}(D_{\mu 3})^{2} \cos\left[2(\rho-\omega)\right]}{M_{2}M_{3}}}} \ast \]
\[ e^{2i\delta_{3}+i\arctan\left[\frac{M_{2}M_{3}(D_{\mu 1})^{2}\sin(2\beta)-M_{1}M_{2}(D_{\mu 3})^{2}\sin(2\rho)+M_{1}M_{3}(D_{\mu 2})^{2}\sin(2\omega)}{M_{2}M_{3}(D_{\mu 1})^{2}\cos(2\beta)-M_{1}M_{2}(D_{\mu 3})^{2}\cos(2\rho)+M_{1}M_{3}(D_{\mu 2})^{2}\cos(2\omega)}\right]}
\]



\begin{thebibliography}{99}

\bibitem{Becquerel}
H. Becquerel, Compt. Ren. 122 (1896) 420; 
H. Becquerel, Compt. Ren. 122 (1896) 501; 
H. Becquerel, Compt. Ren. 122 (1896) 559; 
H. Becquerel, Compt. Ren. 122 (1896) 689; 
H. Becquerel, Compt. Ren. 122 (1896) 762; 
H. Becquerel, Compt. Ren. 122 (1896) 1086; 
H. Becquerel, Compt. Ren. 124 (1897) 438; 
H. Becquerel, Compt. Ren. 124 (1897) 800;
\bibitem{Chadwick}
	J. Chadwick, Verhandl Dtsch. Phys. Ges. 16, 383 (1914).
\bibitem{Ellis}
C. D. Ellis and W. A. Wooster, Proc. Roy. Soc. A117, 109 (1927).
\bibitem{Dirac}
P. A. M. Dirac, Proc. Roy. Soc. A114, 243 (1927); idem, A114, 710
(1927).
\bibitem{Pauli1}
W. Pauli, Open Letter to Radioactive Persons, 1930, for an English
translation see: Physics Today 31, 27 (1978).
\bibitem{Pauli2}
W. Pauli, in Septieme Conseil de Physique, Solvay (Gauthier–
Villars, Paris, 1934), p. 324.
\bibitem{Fermi}
E. Fermi, Z. Phys. 88, 161 (1934); idem, Nuovo Cim. 11, 1 (1934).
\bibitem{Williams}
E. J. Williams and G. E. Roberts, Nature 145, 102 (1940).
\bibitem{Tomonaga}
S. Tomonaga, Bull. I. P. C. R. (Riken Iho) 22, 545 (1943), for the
English translation see Prog. Theor. Phys. 1, 27 (1946).
\bibitem{Feynman}
R. P. Feynman, Phys. Rev. 74, 1430 (1948); idem, 76, 769 (1949).
\bibitem{Schwinger}
J. Schwinger, Phys. Rev. 74, 1439 (1948); idem, 75, 651 (1949).
\bibitem{Tati}
T. Tati and S. Tomonaga, Prog. Theor. Phys. 3, 391 (1948); see also: Z. Koba, T. Tati and S. Tomonaga, Prog. Theor. Phys. 2, 101
and 198 (1947).
\bibitem{Dyson}
F.J. Dyson, Phys. Rev. 75, 486 (1949).
\bibitem{Tiomno}
J. Tiomno and J. A. Wheeler, Rev. Mod. Phys. 21, 144 (1949).
\bibitem{Lee}
T. D. Lee, M. N. Rosenbluth, and C. N. Yang, Phys. Rev. 75, 905 (1949).
\bibitem{Wu}
C. S. Wu, et al., Phys. Rev. 105, 1413 (1957).
\bibitem{Garwin}
R. L. Garwin, L. M. Lederman, and M. Weinrich, Phys. Rev. 105, 1415 (1957).
\bibitem{Friedman}
J. I. Friedman and V. L. Teledgi, Phys. Rev. 105, 1681 (1957).
\bibitem{Kobayashi}
M. Kobayashi and T.Maskawa, Prog. Theor. Phys. 49, 652 (1973).
\bibitem{Schwinger2}
J. Schwinger, Ann. Phys. 2, 407 (1957).
\bibitem{Yang}
T. D. Lee and C. N. Yang, Phys. Rev. 108, 1611 (1957).
\bibitem{Glashow}
S. L. Glashow, Nucl. Phys. 22, 579 (1961).
\bibitem{Hasert}
F. J. Hasert, et al., Phys. Lett. B 46, 121 (1973); idem, B 46, 138(1973).
\bibitem{Arnison}
G. Arnison, et al. (UA1 Collab.), Phys. Lett. B 122, 103 (1983).
\bibitem{Salam}
A. Salam, Nuovo Cim. 5, 299 (1957).
\bibitem{Lee2}
T. D. Lee and C. N. Yang, Phys. Rev. 105, 1671 (1957).
\bibitem{Landau}
L. D. Landau, Zh. Eksp. Teor. Fiz. 32, 407 (1957); see also JETP 5, 337 (1957).
\bibitem{Goldhaber}
M. Goldhaber, L. Grodzins, and A. W. Sunyar, Phys. Rev. 109, 1015, (1958). 
\bibitem{Feynman2}
R. P. Feynman and M. Gell-Mann, Phys. Rev. 109, 193 (1958).
\bibitem{Marshak}
R. E. Marshak and E. C. G. Sudarshan, Phys. Rev. 109, 1860 (1958).
\bibitem{Sakurai}
J. J. Sakurai, Nuovo Cim. 7, 649 (1958).
\bibitem{Theosci}
B. Pontecorvo, JETP, 6, 429 (1958); Z. Maki, Nakagawa and S. Sakata, Prog.
Theor.Phys. 28, 870 (1962).
\bibitem{Goldstone}
J. Goldstone, Nuovo Cim. 19, 154 (1961); see also: Y. Nambu,
Phys. Rev. Lett. 4, 380 (1962).
\bibitem{Higgs}
P. W. Higgs, Phys. Lett. 12, 132 (1964).
\bibitem{Glashow1}
Glashow, S. Gell-Man, M. Gauge Theories of Vector Particles, DOE Technical Report (1961)
\bibitem{SalamWard}
A. Salam and J. C. Ward, Nuovo Cim. 19, 165 (1961).
\bibitem{SalamWard2}
A. Salam and J. C. Ward, Phys. Lett. 13, 168 (1964).
\bibitem{Weinberg2}
S. Weinberg, Phys. Rev. Lett. 19, 1264 (1967).
\bibitem{Abrams}
G. S. Abrams et al. (MARK-II Collab.), Phys. Rev. Lett. 63, 2173 (1989).
\bibitem{SuperK}
Super-Kamiokande Collaboration, Y. Fukuda et al., Phys. Rev. Lett. 81, 1562 (1998); T. Kajita, presented at the ''Neutrino 98'' conference held in Takayama, Japan, 1998.
\bibitem{lep}
Particle Data Gropu. J. Phys. G33, p.478 (2006)
\bibitem{cottin}
Cottingham W.N, \emph{An Introduction to the Standard Model Of Particle Physics}, Cambridge University Press, 1st edition, 2003, p. 104-108, 112-114.
\bibitem{peskin}
Peskin M.E and Schroeder D.V, \emph{An Introduction to Quantum Field Theory}, Westview Press, 1st edition, 1995, p. 690-703.
\bibitem{Weinbergsalam}
S.L. Glashow, Nucl. Phys. B 22, 579 (1961);
S. Weinberg, Phys. Rev. Lett. 19, 1264 (1967);
A. Salam, in: Proceedings of the 8th Nobel Symposium, p. 367, ed. N. Svartholm, Almqvist and Wiksell,
Stockholm 1968
\bibitem{mohabook}
Mohapatra R,
\emph{Massive Neutrinos in Physics and Astrophysics}, World Scientific, 3rd edition, 2004, p. 6.
\bibitem{mohabook2}
Mohapatra R,
\emph{Massive Neutrinos in Physics and Astrophysics}, World Scientific, 3rd edition, 2004, p. 21-25.
\bibitem{bimaximal2}
A. Baltz, A.S. Goldhaber, and M. Goldhaber, Phys. Rev. Lett. 81, 5730 (1998).
\bibitem{bimaximal3}
M. Jezabek and A. Sumino, hep-ph/9807310 (1998).
\bibitem{bimaximal}
V. Barger, S. Pakvasa, T. J. Weiler, and K. Whisnant, hep-ph/ 9806387, (1998). 
\bibitem{maltoni}
M.C. Gonzalez-Garcia, Michele Maltoni , arXiv:0704.1800
\bibitem{osci}
B. T. Cleveland et al., Ap. J. 496, 505 (1998); K. S. Hirata et al., Kamiokande Collaboration,
Phys. Rev. Lett. 77, 1683 (1996); W. Hampel, GALLEX Collaboration, Phys.
Lett. B 388, 384 (1996); J. N. Abdurashitov et al., SAGE Collaboration, Phys. Rev.
Lett. 77, 4708 (1996); Y. Suzuki, Super-Kamiokande Collaboration, Talk presented at
Neutrino '98, Takayama, Japan.
\bibitem{kayser}
Kayser B. et al. \emph{The Physics of Massive Neutrinos}, World Scientific Lecture Notes in Physics Vol. 25, 1989, p. 10-13.
\bibitem{chaichian}
Chaichian M. and Nelipa N.F. \emph{Introduction to Gauge Field Theories}, Springer-Verlag, Germany 1984, p. 153-155.
\bibitem{chau}
L.-L. Chau and W.-Y. Keung, Phys. Rev. Lett. 53, 1802 (1984).
\bibitem{kayser2}
Kayser B. et al. \emph{The Physics of Massive Neutrinos}, World Scientific Lecture Notes in Physics Vol. 25, 1989, p. 77-78.
\bibitem{Bilenky}
S.M. Bilenky. Et al. hep-ph/9812360 (1999)
\bibitem{Bilenky2}
S. M. Bilenky and S. T. Petcov, Rev. Mod. Phys. 59, 671 (1987)
\bibitem{mohabook4}
Mohapatra R,
\emph{Massive Neutrinos in Physics and Astrophysics}, World Scientific, 3rd edition, 2004, p. 21-25.
\bibitem{gonzalez}
M. C. Gonzalez-Garcia And Yosef Nir, Rev. Mod. Phys. 75, 346 (2003)
\bibitem{cheng}
T.P Cheng and Ling-Fong Li, Phys. Rev. D. 22, 2860 (1980)
\bibitem{seesaw}
M. Gell-Mann, P. Ramond and R. Slansky, in Supergravity, edited by P. van Nieuwenhuizen
and D. Freedman, (North-Holland, 1979), p. 315; T. Yanagida, in Proceedings of
the Workshop on the Unified Theory and the Baryon Number in the Universe, edited by
O. Sawada and A. Sugamoto (KEK Report No. 79-18, Tsukuba, 1979), p. 95; R.N. Mohapatra
and G. Senjanovi´c, Phys. Rev. Lett. 44 (1980) 912.
\bibitem{mohabook3}
Mohapatra R, \emph{Massive Neutrinos in Physics and Astrophysics}, World Scientific, 3rd edition, 2004, p. 141-143. also check Kayser B. et al. \emph{The Physics of Massive Neutrinos}, World Scientific Lecture Notes in Physics Vol. 25, 1989, p. 84-92.
\bibitem{Spergel}
D. N. Spergel et al., astro-ph/0603449.
\bibitem{Sakharov}
A. D. Sakharov, JETP Lett. 5 (1967) 24.
\bibitem{Fukugita}
M. Fukugita and T. Yanagida, Phys. Lett. B174 (1986) 45.
\bibitem{Hooft}
G. 't Hooft, Phys. Rev. Lett. 37, 8 (1976).
\bibitem{KRS}
Kuzmin, V.A. et al. PHYS. Lett. 155, 36 (1985).
\bibitem{cohen}
For review see, Cohen, A.G, et al. Ann. Rev. Nucl. Part. Sci. 43, 27 (1983).
\bibitem{Avignone}
Avignone III, F. arXiv:0708.1033v1
\bibitem{juan}
Juan Carlos G\'omez-Izquierdo, Abdel P\'erez-Lorenzana, arXiv:0711.0045
\end{thebibliography}
\end{document}